\documentclass[12pt,english]{article}
\usepackage{palatino}
\usepackage[latin1]{inputenc}
\usepackage{a4wide}
\usepackage{graphicx}
\usepackage{amssymb}

\makeatletter

\providecommand{\tabularnewline}{\\}

 \usepackage{amsthm, amsmath, amsfonts, amssymb}
 \theoremstyle{plain}    
 \newtheorem{thm}{Theorem}
 \theoremstyle{definition}
 \newtheorem{defn}[thm]{Definition}
 \theoremstyle{plain}    
 \newtheorem{lem}[thm]{Lemma} 
 \theoremstyle{plain}    
 \newtheorem{prop}[thm]{Proposition} 

 \theoremstyle{definition}
  \newtheorem{example}[thm]{Example}

\usepackage{multicol}
\usepackage{floatflt}
\usepackage{graphicx}

\usepackage{babel}
\makeatother
\begin{document}

\title{Intersections of Lagrangian submanifolds and the Mel'nikov $1$-form}

\author{Nicolas Roy%
\thanks{We are very grateful to Y. Colin de Verdière who pointed this issue
out to us.%
}~%
\thanks{We would like to thank the referee for his careful reading of the
manuscript and for pointing out several inaccuracies and missing bibliographic
references.%
}}

\maketitle
\begin{abstract}
We make explicit the geometric content of Mel'nikov's method for detecting
heteroclinic points between transversally hyperbolic periodic orbits.
After developing the general theory of intersections for pairs of
families of Lagrangian submanifolds $\mathcal{N}_{\varepsilon}^{\pm}$,
with $\mathcal{N}_{0}^{+}=\mathcal{N}_{0}^{-}$ and constrained to
live in an auxiliary family of submanifolds, we explain how the heteroclinic
orbits of a given Hamiltonian system are detected by the zeros of
the Mel'nikov $1$-form. This $1$-form admits an integral expression
which is non-convergent in general. We discuss different solutions
to this convergence problem.
\end{abstract}

\section*{Introduction }

In his article \cite{melnikov}, Mel'nikov introduced a method for
studying time-periodic perturbations $H_{\varepsilon}\left(x,\xi,t\right)=H_{0}\left(x,\xi\right)+\varepsilon H_{1}\left(x,\xi,t\right)$
of $2$-dimensional time-independent Hamiltonian systems. The author
considers the case where $H_{0}$ has a hyperbolic fixed point $m_{0}\in\mathbb{R}^{2}$
such that (one {}``half'' of) its stable manifold coincides with
(one {}``half'' of) its unstable manifold, as depicted on the picture
below. \begin{floatingfigure}[l]{4.5cm}\includegraphics[%
  width=4cm]{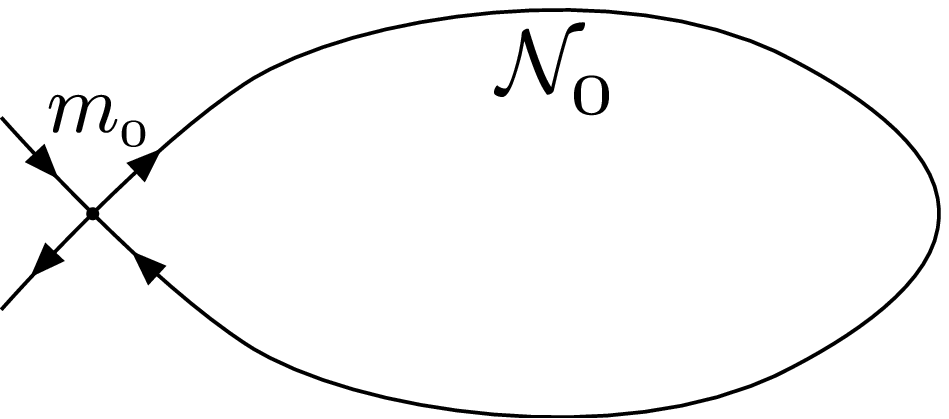}

\includegraphics[%
  width=4cm]{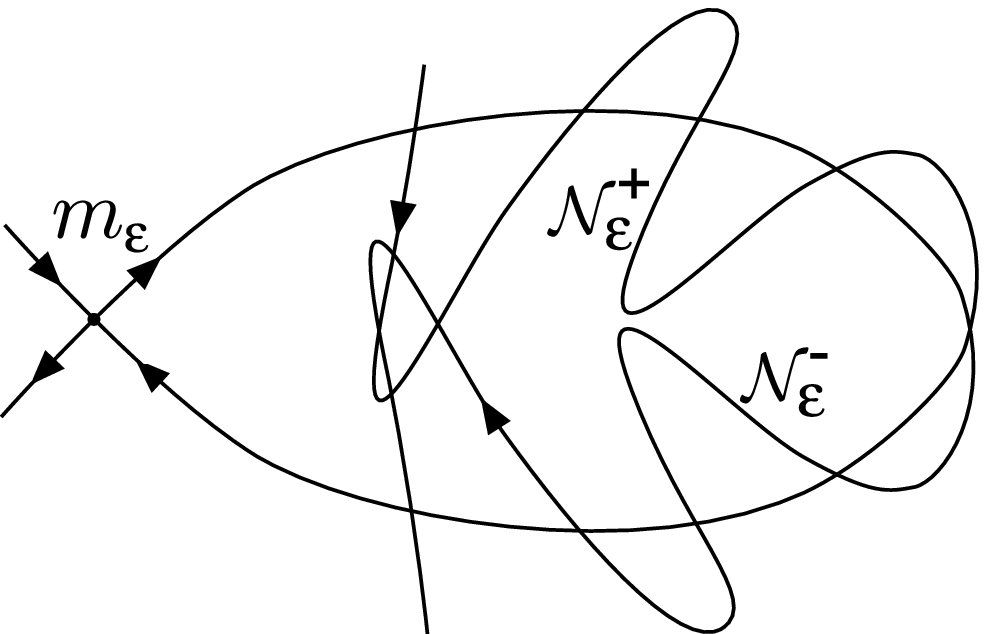}\end{floatingfigure} 

Let us denote this manifold by $\mathcal{N}_{0}$. For studying the
time-dependent perturbations of $H_{0}$, one might consider a section
at time $t=0$ of the system in $\mathbb{R}^{2}\times S^{1}$, given
by the time $1$ flow $\phi_{X_{H_{\varepsilon}}}^{t=1}$. Because
of the structural stability of hyperbolic points, there is a smooth
family $m_{\varepsilon}$ of hyperbolic points for the map $\phi_{X_{H_{\varepsilon}}}^{1}$.
Furthermore, the hyperbolicity implies the existence of a smooth family
of stable (resp. unstable) manifolds $\mathcal{N}_{\varepsilon}^{+}$
(resp. $\mathcal{N}_{\varepsilon}^{-}$) for $m_{\varepsilon}$. However,
as soon as $\varepsilon\neq0$, they might not coincide and their
intersections (called homoclinic points) form in general a very complicated
set. See the picture on the left. 

This phenomenon, referred to as the {}``homoclinic entanglement'',
is the sign of the chaotic behaviour of the system near $m_{0}$.
It is also known to be the key feature of Arnold's diffusion (see
e.g. \cite{chierchia_gallavotti}). In order to detect the positions
of the homoclinic points of $m_{\varepsilon}$, Mel'nikov defined
the function \[
M\left(t\right)=\int_{-\infty}^{+\infty}\left\{ H_{1}\left(t+s\right),H_{0}\right\} \left(m\left(s\right)\right)\, ds,\]
 where $m\left(s\right)$ is the trajectory on $\mathcal{N}_{0}$
under the unperturbed dynamics of $H_{0}$ starting from a chosen
point $m\in\mathcal{N}_{0}$. This point plays the role of an origin
on $\mathcal{N}_{0}$ and $t$ is a coordinate. Mel'nikov shows that
the non-degenerate zeros of $M$ describe at first order in $\varepsilon$
the position of the homoclinic points of the perturbed hyperbolic
point $m_{\varepsilon}$. The main feature of the expression of $M$
is that the only flow that one has to integrate is the one of $H_{0}$,
i.e.,, the unperturbed dynamics, which is supposed to be well understood.
On the other hand, one knows that such a time periodic perturbed system
can be rewritten as an autonomous one, thru a standard procedure.
Namely, one takes the product of the initial symplectic manifold (here
simply $\mathbb{R}^{2}$) with $T^{*}S^{1}$, where the $S^{1}$ factor
corresponds to the $t$ variable. In the extended system, the hyperbolic
fixed point $m_{\varepsilon}$ becomes a transversally hyperbolic
periodic orbit $\gamma_{\varepsilon}$, whose stable and unstable
manifolds intersect along trajectories homoclinic to $\gamma_{\varepsilon}$. 

The main goal of this article is to clarify the geometric content
of Mel'ni{\-}kov's method, which extends to higher-dimensional systems
on general symplectic manifolds, for detecting heteroclinic (and not
only homoclinic) orbits linking two periodic orbits. The Mel'ni{\-}kov's
method has actually two separate aspects. First, the heteroclinic
orbits are in correspondence with the zeros of a geometric object,
namely the Mel'ni{\-}kov's $1$-form. Second, one tries to give this
$1$-form an integral expression involving only the flow of the unperturbed
dynamics. These two issues roughly correspond to the main two sections
of this paper.

The extension of Mel'ni{\-}kov's technique for detecting heteroclinic
orbits linking two transversally hyperbolic periodic orbits or tori,
rather than hyperbolic points, has been considered by many authors,
e.g. \cite{arnold_7,delshams_guti,delshams_guti_2,eliasson_3,holmes_marsden,lochak_marco_sauzin,wiggins}.
But, they all consider dynamical systems with the common feature that
there is an explicit separation between the {}``longitudinal'' and
{}``transversal'' variables, corresponding respectively to the motion
along the tori (or the periodic orbits) and the hyperbolic transversal
motion. It turns out that resorting to explicit coordinates has several
drawbacks we would like to point out now.

\begin{itemize}
\item First of all, this assumption is unnecessary and actually goes against
a satisfactory understanding of the geometry underlying this method.
One aim of this paper is to describe the geometric objects involved
in Mel'ni{\-}kov's method without reference to any coordinate system.
In particular, as a multidimensional generalisation of the Mel'nikov
function, the authors introduce a {}``Mel'nikov vector'', whereas
the correct geometric object is rather a $1$-form, as we explain
throughout this paper. The use of a $1$-form is in fact very natural
since Mel'ni{\-}kov's method deals with deformations of Lagrangian
submanifolds (the stable and unstable ones) and it is well-known \cite{weinstein1}
that the deformation theory of Lagrangian submanifolds is parameterised
by closed $1$-forms. The Mel'ni{\-}kov's $1$-form is thus closed
and it is actually exact for geometrical reasons explained in Section
\ref{sub_Melnikov_potentials}. We believe that this clarifies the
statement {}``The Mel'ni{\-}kov vector is a gradient'' which, in
the literature, seems to be true for a bit obscure reasons. In fact,
this is always true and not only in the particular models people studied.
\item Second, these particular models (with a separation between the longitudinal
and transversal motions) dismiss a large class of systems. Indeed,
it is well-known from different studies of completely integrable systems
\cite{colin_vungoc,zung_6} that the local model near a transversally
hyperbolic invariant $m$-dimensional torus is not always $\mathbb{T}^{m}\times\mathbb{R}^{m}\times\mathbb{R}^{2n}$
(as in the mentioned particular models) but may be a quotient of that
by a finite group. For example, in dimension $4$, it happens that
the local stable and unstable manifolds of a periodic orbit is not
diffeomorphic to the cylinder but rather to the Möbius strip%
\footnote{See \cite{colin_vungoc} for a precise description of such systems.
See also the end of section \ref{sub_moebius} for a picture of this
situation.%
}. 
\item Third, these particular systems are highly non-generic in the heteroclinic
case. Indeed, they have the feature that the heteroclinic manifolds
link two tori with the same Diophantine property. For example, in
the case of periodic orbits (instead of tori), this means that the
orbits have the same period. Generically, the periods are different
and this prevents us from expressing the Mel'ni{\-}kov $1$-form
in terms of an integral over the unperturbed flow. This issue is treated
in Section \ref{sec_good_A}.
\end{itemize}
The general tool we will rely on is the intersection theory for pairs
$\left(\mathcal{N}_{\varepsilon}^{+},\mathcal{N}_{\varepsilon}^{-}\right)$
of Lagrangian submanifolds which coincide for $\varepsilon=0$ and
which are constrained to live in an auxiliary submanifold $\mathcal{N}_{\varepsilon}^{\pm}\subset\mathcal{P}_{\varepsilon}$
for all $\varepsilon$. Indeed, stable and unstable manifolds of transversally
hyperbolic periodic orbits are Lagrangian and confined at least in
an energy level $\left\{ H_{\varepsilon}=cst\right\} $. For this
particular intersection theory, one has to introduce a suitable {}``transversality''
condition at $\varepsilon=0$ (roughly speaking, a condition on the
variations {}``$\frac{d\mathcal{N}_{\varepsilon}^{\pm}}{d\varepsilon}$'')
in order to insure transversality of $\mathcal{N}_{\varepsilon}^{+}$
and $\mathcal{N}_{\varepsilon}^{-}$ in \emph{$\mathcal{P}_{\varepsilon}$}
for $\varepsilon\neq0$, since the usual transversality hypothesis
is obviously not fulfilled at $\varepsilon=0$. This theory, which
actually applies to any pair of Lagrangian submanifolds regardless
to their stable/unstable feature, is developed in Section \ref{sec_intersections}.
It is shown that investigating the intersections of $\mathcal{N}_{\varepsilon}^{+}$
and $\mathcal{N}_{\varepsilon}^{-}$ for $\varepsilon\neq0$ amounts
to looking for the {}``non-degenerate'' zeros of a $1$-form $\beta$
defined on $\mathcal{N}_{0}^{+}=\mathcal{N}_{0}^{-}$, which we call
the \emph{Mel'nikov $1$-form} despite this name takes on its full
meaning only when $\mathcal{N}_{\varepsilon}^{+}$ and $\mathcal{N}_{\varepsilon}^{-}$
are the stable and unstable manifolds of two transversally hyperbolic
periodic orbits $\gamma_{\varepsilon}^{\pm}$. In that case, the intersections
of $\mathcal{N}_{\varepsilon}^{+}$ and $\mathcal{N}_{\varepsilon}^{-}$
are thus heteroclinic points between $\gamma_{\varepsilon}^{+}$ and
$\gamma_{\varepsilon}^{-}$, or homoclinic points in case there is
only one periodic orbit $\gamma_{\varepsilon}^{+}=\gamma_{\varepsilon}^{-}$. 

This is the topic of Section \ref{sec_melnikov}, where we apply the
theory developed in Section \ref{sec_intersections} to this heteroclinic/homoclinic
situation. We will focus on the following questions. When the unperturbed
Hamiltonian $H_{0}$ is completely integrable (this is automatic for
$2$-dimensional systems), i.e., admits a momentum map $\mathbf{A}=\left(A_{1},...,A_{d}\right)$,
then one can compute the Mel'nikov $1$-form $\beta$ thru the evaluations
$\beta\left(X_{A_{j}}\right)$. This shows in particular that in the
near-integrable case, the splitting of the stable and unstable manifolds
is completely described by the integrals of motion of the unperturbed
Hamiltonian $H_{0}$. Beside this, it turns out that the functions
$\beta\left(X_{A_{j}}\right)$ have an integral expression involving
only the flow of $H_{0}$. Unfortunately, these integrals do not converge
in general. Then, we discuss what are the different solutions to this
convergence problem, namely either assuming that the perturbation
$H_{\varepsilon}-H_{0}$ is critical on both orbits $\gamma_{0}^{\pm}$
or choosing the $A_{j}$'s critical on $\gamma_{0}^{\pm}$. The latter
works perfectly in the homoclinic situation, but we explain that in
the heteroclinic one, there is usually not enough independent such
$A_{j}$'s to determine the Mel'nikov $1$-form. We show however that
there is a special case (to which belong the time-periodic systems)
for which there is precisely enough $A_{j}$'s critical on $\gamma_{0}^{\pm}$
to compute $\beta$. This question is usually ignored in the literature
since the authors consider either the homoclinic situation or periodically
forced systems.

\section{Intersections of families of Lagrangian submanifolds\label{sec_intersections}}

We forget for the moment the heteroclinic theory of transversally
hyperbolic orbits and we begin with the intersection theory for some
families of compact submanifolds $\mathcal{N}_{\varepsilon}^{\pm}$
in a given manifold $\mathcal{M}$. All the manifolds under consideration
are smooth. As well, we assume that the families depend smoothly on
the deformation parameter $\varepsilon$, in the sense that the union
$\bigcup_{\varepsilon}\left(\mathcal{N}_{\varepsilon}^{\pm}\times\left\{ \varepsilon\right\} \right)$
is a smooth submanifold of $\mathcal{M}\times\mathbb{R}$. From now
on, both these smoothness conditions will always be implicitly assumed.

It is well-known that whenever $\mathcal{N}_{0}^{+}$ and $\mathcal{N}_{0}^{-}$
intersect transversally at some point $m$, i.e., $T_{m}\mathcal{N}_{0}^{+}\oplus T_{m}\mathcal{N}_{0}^{-}=T_{m}\mathcal{M}$,
then in a neighbourhood of $m$, the intersection $\mathcal{N}_{0}^{+}\cap\mathcal{N}_{0}^{-}$
is a smooth submanifold of dimension equal to $\dim\mathcal{N}_{0}^{+}+\dim\mathcal{N}_{0}^{-}-\dim\mathcal{M}$.
Moreover, $\mathcal{N}_{\varepsilon}^{+}\cap\mathcal{N}_{\varepsilon}^{-}$
is a smooth family of submanifolds of $\mathcal{M}$, for small enough
$\varepsilon$. 

\begin{floatingfigure}[l]{6cm}\includegraphics[%
  scale=0.5]{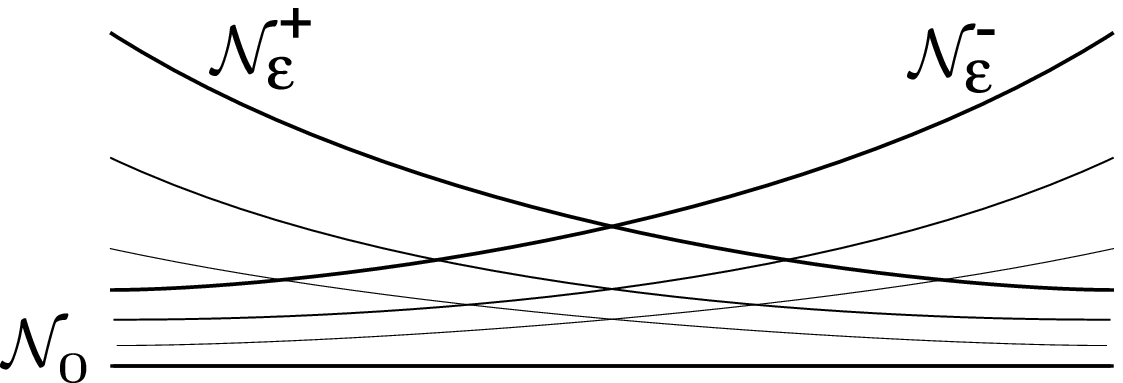}\end{floatingfigure}As mentioned in the introduction, we need to
consider the situation where $\mathcal{N}_{\varepsilon}^{+}$ and
$\mathcal{N}_{\varepsilon}^{-}$ are deformations of the same $\mathcal{N}_{0}$.
Such families are obviously not transverse for $\varepsilon=0$, but
a suitable transversality condition on the {}``first derivatives
$\frac{d}{d\varepsilon}\mathcal{N}_{\varepsilon}^{\pm}$'' can be
introduced to describe the intersection $\mathcal{N}_{\varepsilon}^{+}\cap\mathcal{N}_{\varepsilon}^{-}$
for $\varepsilon\neq0$. This issue is addressed in Section \ref{sec_infinitesimal_intersection}.

On the other hand, if we know a priori that $\mathcal{N}_{\varepsilon}^{+}$
and $\mathcal{N}_{\varepsilon}^{-}$ are constrained to live in an
intermediate submanifold $\mathcal{P}_{\varepsilon}$, the smoothness
of the intersection $\mathcal{N}_{\varepsilon}^{+}\cap\mathcal{N}_{\varepsilon}^{-}$
can be insured by a {}``infinitesimal transversality'' condition
in $\mathcal{P}_{0}$. This is precisely the case for the Mel'nikov
situation where the families under consideration are included in a
level set of the Hamiltonian function $H_{\varepsilon}$. This question
is considered in Section \ref{sec_constraint}.

Eventually, in the symplectic framework, the intersection theory for
Lagrangian submanifolds is somehow simpler and it is well-described
by the Mel'nikov $1$-form, a differential form on $\mathcal{N}_{0}$,
as we discuss in Section \ref{sec_lagrangian_intersection}.

\subsection{Infinitesimally transverse intersections\label{sec_infinitesimal_intersection}}

\subsubsection{Generating flows for families of submanifolds }

First, we need to parameterise the families of submanifolds with families
of diffeomorphisms in the following way.

\begin{defn}
\label{def_generating_flow}Let $\mathcal{N}_{\varepsilon}\subset\mathcal{M}$
be a family of compact submanifolds. A (time-dependent) vector field
$X_{\varepsilon}$ is said to \textbf{generate} $\mathcal{N}_{\varepsilon}$
if its flow $\phi_{X_{\varepsilon}}^{\varepsilon}$ satisfies $\phi_{X_{\varepsilon}}^{\varepsilon}\left(\mathcal{N}_{0}\right)=\mathcal{N}_{\varepsilon}$
and if $X_{0}$ is not tangent to $\mathcal{N}_{0}$ whenever it is
non-zero, i.e.,, \[
X_{0}\left(m\right)\in T_{m}\mathcal{N}_{0}\Longrightarrow X_{0}\left(m\right)=0.\]
We will also say that the flow $\phi_{X_{\varepsilon}}^{\varepsilon}$
generates $\mathcal{N}_{\varepsilon}$.
\end{defn}
Notice that in general it is impossible to a choose a time-independent
vector field to generate a given family $\mathcal{N}_{\varepsilon}$,
whereas there always exists a time-dependent one, as the next lemma
shows.

\begin{lem}
\label{lem_generating_flow_exist}Let $\mathcal{N}_{\varepsilon}\subset\mathcal{M}$
be a family of compact submanifolds. Then there exists a vector field
$X_{\varepsilon}$ generating $\mathcal{N}_{\varepsilon}$. Moreover,
when one is given two deformations $\mathcal{N}_{\varepsilon}^{\pm}$
of the same $\mathcal{N}_{0}:=\mathcal{N}_{0}^{+}=\mathcal{N}_{0}^{-}$,
then there exist generating vector fields $X_{\varepsilon}^{\pm}$
such that $X_{0}^{+}-X_{0}^{-}$ is not tangent to $\mathcal{N}_{0}$
whenever it is non-zero.
\end{lem}
\begin{proof}
The Tubular Neighbourhood Theorem says that there is a neighborhood
$\mathcal{O}\subset\mathcal{M}$ of $\mathcal{N}_{0}$, a vector bundle
$E$ over $\mathcal{N}_{0}$, a neighborhood $\tilde{\mathcal{O}}\subset E$
of the zero-section and a diffeomorphism $\chi:\mathcal{O}\rightarrow\tilde{\mathcal{O}}$
which sends $\mathcal{N}_{0}$ to the zero-section. One can assume
$\mathcal{O}$ and $\tilde{\mathcal{O}}$ are compact. For small enough
$\varepsilon$, the submanifolds $\mathcal{N}_{\varepsilon}^{\pm}$
lie in $\mathcal{O}$. Thru the map $\chi$, one obtains families
of submanifolds $\tilde{\mathcal{N}}_{\varepsilon}^{\pm}=\chi\left(\mathcal{N}_{\varepsilon}^{\pm}\right)\subset\tilde{\mathcal{O}}$
close to the zero-section, hence they are graphs of sections of $E$.
Therefore, there exist particular vector fields $\tilde{Y}_{\varepsilon}^{\pm}$
generating $\tilde{\mathcal{N}}_{\varepsilon}^{\pm}$, namely those
associated to vertical translations. These vector fields are vertical
and thus not tangent to $\mathcal{N}_{0}$ whenever they do not vanish.
The same is true for the difference $Y_{\varepsilon}^{+}-Y_{\varepsilon}^{-}$.
Now, define $\tilde{X}_{\varepsilon}^{\pm}=f\tilde{Y}_{\varepsilon}^{\pm}$,
with $f\in C_{0}^{\infty}\left(E\right)$ a smooth function with support
in $\tilde{\mathcal{O}}$ and equal to $1$ in a (smaller) neighborhood
of the zero section. By construction, for small enough $\varepsilon$,
the vector fields $X_{\varepsilon}^{\pm}=\chi_{*}^{-1}\left(\tilde{X}_{\varepsilon}^{\pm}\right)$
on $\mathcal{M}$ generate $\mathcal{N}_{\varepsilon}^{\pm}$ in the
sense of Definition \ref{def_generating_flow}.
\end{proof}
In all the following, we will always choose implicitly generating
vector fields with the property of Lemma \ref{lem_generating_flow_exist}.
For a given family $\mathcal{N}_{\varepsilon}$, the choice of a generating
vector field is of course not unique, but different choices are related
as follows.

\begin{lem}
\label{lem_generating_flow_non_unique}Let $\mathcal{N}_{\varepsilon}\subset\mathcal{M}$
be a family of compact submanifolds. If two vector fields $X_{\varepsilon}$
and $Y_{\varepsilon}$ generate $\mathcal{N}_{\varepsilon}$, then
the difference $X_{\varepsilon}-Y_{\varepsilon}$ is tangent to $\mathcal{N}_{\varepsilon}$
for all $\varepsilon$.
\end{lem}
\begin{proof}
Denote by $\phi^{\varepsilon}$ (resp. $\varphi^{\varepsilon}$) the
flow of $X_{\varepsilon}$ (resp. $Y_{\varepsilon}$). The vector
field of the flow $\left(\phi^{\varepsilon}\right)^{-1}$ is equal
to $-\left(\phi^{\varepsilon}\right)_{*}^{-1}\left(X_{\varepsilon}\right)$
and therefore, the composition $\psi^{\varepsilon}=\left(\phi^{\varepsilon}\right)^{-1}\circ\varphi^{\varepsilon}$
is the flow of the vector field $\left(\phi^{\varepsilon}\right)_{*}^{-1}\left(-X_{\varepsilon}+Y_{\varepsilon}\right)$.
On the other hand, $\psi^{\varepsilon}$ obviously sends $\mathcal{N}_{0}$
to itself and its vector field is thus tangent to $\mathcal{N}_{0}$.
This implies that $X_{\varepsilon}-Y_{\varepsilon}$ is tangent to
$\phi^{\varepsilon}\left(\mathcal{N}_{0}\right)=\mathcal{N}_{\varepsilon}$,
for all $\varepsilon$.
\end{proof}

\subsubsection{Infinitesimal transversality}

From now on, we consider two families $\mathcal{N}_{\varepsilon}^{+}$
and $\mathcal{N}_{\varepsilon}^{-}$ which are deformations of the
same submanifold $\mathcal{N}_{0}:=\mathcal{N}_{0}^{+}=\mathcal{N}_{0}^{-}$.
Let us now introduce the suitable transversality condition to describe
the intersections of $\mathcal{N}_{\varepsilon}^{+}$ and $\mathcal{N}_{\varepsilon}^{-}$.

\begin{defn}
\label{def_infinit_intersec}Let $\mathcal{N}_{\varepsilon}^{\pm}\subset\mathcal{M}$
be two families of compact submanifolds generated by vector fields
$X_{\varepsilon}^{\pm}$. We say that a point $m\in\mathcal{N}_{0}$
is an \textbf{infinitesimal intersection of $\mathcal{N}_{\varepsilon}^{+}$
and $\mathcal{N}_{\varepsilon}^{-}$} if $X_{0}^{+}=X_{0}^{-}$ at
$m$. 
\end{defn}
\begin{lem}
\label{lem_infinit_intersec_bien_def}The notion of {}``infinitesimal
intersection'' is well-defined, i.e., independent of the choice of
$X_{\varepsilon}^{\pm}$.
\end{lem}
\begin{proof}
Let $X_{\varepsilon}^{+}$ (resp. $X_{\varepsilon}^{-}$) be a vector
field generating $\mathcal{N}_{\varepsilon}^{+}$ (resp. $\mathcal{N}_{\varepsilon}^{-}$)
and let $m\in\mathcal{N}_{0}$ be a point where $X_{0}^{+}=X_{0}^{-}$.
Suppose we have a second vector field $\tilde{X}_{\varepsilon}^{+}$
(resp. $\tilde{X}_{\varepsilon}^{-}$) generating $\mathcal{N}_{\varepsilon}^{+}$
(resp. $\mathcal{N}_{\varepsilon}^{-}$). According to Lemma \ref{lem_generating_flow_non_unique},
the differences $X_{0}^{+}-\tilde{X}_{0}^{+}$ and $X_{0}^{-}-\tilde{X}_{0}^{-}$
are tangent to $\mathcal{N}_{0}$ and therefore so is $X_{0}^{+}-X_{0}^{-}-\left(\tilde{X}_{0}^{+}-\tilde{X}_{0}^{-}\right)$.
Now, if $X_{0}^{+}-X_{0}^{-}$ vanishes at $m$, then $\tilde{X}_{0}^{+}-\tilde{X}_{0}^{-}$
must be tangent to $\mathcal{N}_{0}$. This is a contradiction and
therefore $\tilde{X}_{0}^{+}-\tilde{X}_{0}^{-}$ vanishes at $m$
too.
\end{proof}
\begin{defn}
\label{def_operator_D}Let $X_{\varepsilon}^{\pm}$ be vector fields
generating $\mathcal{N}_{\varepsilon}^{\pm}$ and let $m\in\mathcal{N}_{0}$
be an infinitesimal intersection of $\mathcal{N}_{\varepsilon}^{+}$
and $\mathcal{N}_{\varepsilon}^{-}$. We define the linear operator
$D_{m,X_{0}^{\pm}}:T_{m}\mathcal{N}_{0}\rightarrow T_{m}\mathcal{M}$
by \[
D_{m,X_{0}^{\pm}}\left(Y\right):=\left[\tilde{Y},X_{0}^{+}-X_{0}^{-}\right]_{m},\]
where $\tilde{Y}\in\Gamma\left(T\mathcal{M}\right)$ is any extension
to $\mathcal{M}$ of $\iota_{*}Y$, with $\iota:\mathcal{N}_{0}\hookrightarrow\mathcal{M}$
the inclusion map. 
\end{defn}
\begin{lem}
\label{lem_op_D_bien_def}The operator $D_{m,X_{0}^{\pm}}$ is well-defined,
i.e., independent of the choice of the extension $\tilde{Y}$. 
\end{lem}
\begin{proof}
Let $Y\in T_{m}\mathcal{N}_{0}$ be a vector. If $\tilde{Y}$ and
$\tilde{Y}^{'}$ are two extensions of $\iota_{*}Y$, then the difference
$\tilde{Y}^{'}-\tilde{Y}$ vanishes at $m$ and we have $\left[\tilde{Y}^{'}-\tilde{Y},X_{0}^{+}-X_{0}^{-}\right]_{m}=0$
since $X_{0}^{+}-X_{0}^{-}$ also vanishes at $m$. The definition
of $D_{m,X_{0}^{\pm}}$ is thus independent of the choice of the extension
$\tilde{Y}$. 
\end{proof}
Notice that despite the operator $D_{m,X_{0}^{\pm}}$ depends on the
choice of the generating vector fields $X_{\varepsilon}^{\pm}$, the
following notion does not. 

\begin{defn}
\label{def_transverse}Let $\mathcal{N}_{\varepsilon}^{\pm}$ be families
of compact submanifolds. An infinitesimal intersection $m\in\mathcal{N}_{0}$
is called a \textbf{transverse} whenever the space $\textrm{img}D_{m,X_{0}^{\pm}}$
is transverse to $T_{m}\mathcal{N}_{0}$ in $T_{m}\mathcal{M}$, with
$X_{\varepsilon}^{\pm}$ any generating vector fields.
\end{defn}
\begin{lem}
\label{lem_transverse_bien_def}The previous notion of transversality
is well-defined, i.e., independent of the choice of the vector fields
generating $\mathcal{N}_{\varepsilon}^{\pm}$.
\end{lem}
\begin{proof}
Suppose we have two pairs of vector fields $X_{\varepsilon}^{\pm}$
and $\tilde{X}_{\varepsilon}^{\pm}$ generating $\mathcal{N}_{\varepsilon}^{\pm}$.
The operators $D_{m,\tilde{X}_{0}^{\pm}}$ and $D_{m,X_{0}^{\pm}}$
are then simply related by \[
D_{m,\tilde{X}_{0}^{\pm}}Y=D_{m,X_{0}^{\pm}}Y+\left[\tilde{Y},X_{0}^{+}-X_{0}^{-}-\left(\tilde{X}_{0}^{+}-\tilde{X}_{0}^{-}\right)\right]_{m}.\]
According to Lemma \ref{lem_generating_flow_non_unique}, both differences
$X_{0}^{+}-\tilde{X}_{0}^{+}$ and $X_{0}^{-}-\tilde{X}_{0}^{-}$
are tangent to $\mathcal{N}_{0}$ and therefore so is $X_{0}^{+}-X_{0}^{-}-\left(\tilde{X}_{0}^{+}-\tilde{X}_{0}^{-}\right)$.
Since $D_{m,\tilde{X}_{0}^{\pm}}\left(Y\right)$ and $D_{m,X_{0}^{\pm}}\left(Y\right)$
are independent of the choice of the extension $\tilde{Y}$, we can
choose it to be tangent to $\mathcal{N}_{0}$. Therefore, the Lie
bracket is also tangent to $\mathcal{N}_{0}$. This implies that $\textrm{img}D_{m,X_{0}^{\pm}}$
is transverse to $T_{m}\mathcal{N}_{0}$ iff $\textrm{img}D_{m,\tilde{X}_{0}^{\pm}}$
is. 
\end{proof}
We give now an equivalent and convenient criterion for the infinitesimal
transversality.

\begin{lem}
\label{lem_transverse_intersect_dim_noyau}Let $\mathcal{N}_{\varepsilon}^{\pm}$
be families of compact submanifolds. There exist generating vector
fields $X_{\varepsilon}^{\pm}$ such that for any infinitesimal intersection
$m\in\mathcal{N}_{0}$, the space $\textrm{img}\left(D_{m,X_{0}^{\pm}}\right)$
does not intersect $T_{m}\mathcal{N}_{0}$. For such $X_{\varepsilon}^{\pm}$,
$m$ is transverse iff\[
\dim\left(\textrm{ker}D_{m,X_{0}^{\pm}}\right)=2\dim\mathcal{N}_{0}-\dim\mathcal{M}.\]

\end{lem}
\begin{proof}
First, proceeding as in the proof of Lemma \ref{lem_generating_flow_exist},
we reduce to families of submanifolds in a neighbourhood $\mathcal{O}$
of the zero-section of a vector bundle over $\mathcal{N}_{0}$, and
we can choose the generating vector fields to be vertical translations.
Moreover, if for the evaluation $D_{m,X_{0}^{\pm}}\left(Y\right):=\left[\tilde{Y},X_{0}^{+}-X_{0}^{-}\right]_{m}$
we choose an extension $\tilde{Y}$ which is a lift of a vector field
on $\mathcal{N}_{0}$, then the Lie bracket is vertical. This implies
that the intersection $\textrm{img}\left(D_{m,X_{0}^{\pm}}\right)\cap T_{m}\mathcal{N}_{0}$
reduces to $\left\{ 0\right\} $. For the second point, we notice
that the vector spaces $\textrm{img}\left(D_{m,X_{0}^{\pm}}\right)$
and $T_{m}\mathcal{N}_{0}$ are transverse in $T_{m}\mathcal{M}$
iff the dimension of the intersection $\textrm{img}\left(D_{m,X_{0}^{\pm}}\right)\cap T_{m}\mathcal{N}_{0}$
is equal to $\dim\left(\textrm{img}D_{m,X_{0}^{\pm}}\right)+\dim T_{m}\mathcal{N}_{0}-\dim T_{m}\mathcal{M}$.
Since the intersection is $\left\{ 0\right\} $, the transversality
condition amounts to requiring that $\dim\left(\textrm{img}D_{m,X_{0}^{\pm}}\right)+\dim T_{m}\mathcal{N}_{0}-\dim T_{m}\mathcal{M}=0$.
Using then the fact that $\dim\left(\textrm{img}D_{m,X_{0}^{\pm}}\right)=\dim\mathcal{N}_{0}-\dim\left(\ker D_{m,X_{0}^{\pm}}\right)$,
we obtain the claimed expression.
\end{proof}
We now state the theorem which shows that the infinitesimal transversality
is the good notion for our problem.

\begin{thm}
\label{theo_2manifold_transverse_intersect}Let $\mathcal{N}_{\varepsilon}^{\pm}\subset\mathcal{M}$
be families of compact submanifolds. If $m\in\mathcal{N}_{0}$ is
a transverse infinitesimal intersection, then near $m$ there is a
smooth family of submanifolds $\Lambda_{\varepsilon}$ with $\Lambda_{0}\subset\mathcal{N}_{0}$
and $\Lambda_{\varepsilon}=\mathcal{N}_{\varepsilon}^{+}\cap\mathcal{N}_{\varepsilon}^{-}$
for small enough $\varepsilon\neq0$.
\end{thm}
\begin{proof}
The proof consists of four arguments.
\begin{itemize}
\item First, applying the Tubular Neighbourhood for $\mathcal{N}_{0}$,
we transpose the situation to a compact neighbourhood of the zero-section
of a vector bundle $E$ over $\mathcal{N}_{0}$. Denote by $\pi:E\rightarrow\mathcal{N}_{0}$
the projection and $\iota:\mathcal{N}_{0}\rightarrow E$ the inclusion
map. For small enough $\varepsilon$, the manifolds $\mathcal{N}_{\varepsilon}^{\pm}$
are the graphs of sections, denoted by $\alpha_{\varepsilon}^{\pm}:\mathcal{N}_{0}\rightarrow E$,
with $\pi\circ\alpha_{\varepsilon}^{\pm}=\mathbb{I}_{\mathcal{N}_{0}}$,
which satisfy $\alpha_{0}^{\pm}=0$. Then, we choose generating vector
fields $X_{\varepsilon}^{\pm}$ which are vertical and constant on
the fibers. In other words, we have $X_{\varepsilon}^{\pm}=\frac{d\alpha_{\varepsilon}^{\pm}}{d\varepsilon}$
if we identify the fibres with their tangent space.
\item Second, denote by $\mathcal{N}_{\varepsilon}=\left(\phi_{X_{\varepsilon}^{-}}^{\varepsilon}\right)^{-1}\left(\mathcal{N}_{\varepsilon}^{+}\right)$
and $\alpha_{\varepsilon}:\mathcal{N}_{0}\rightarrow E$ the associated
family of sections. One checks easily that $\mathcal{N}_{\varepsilon}^{+}$
and $\mathcal{N}_{\varepsilon}^{-}$ intersect transversally at a
point $m$ iff $\mathcal{N}_{\varepsilon}$ and $\mathcal{N}_{0}$
do at the point $\left(\phi_{X_{\varepsilon}^{-}}^{\varepsilon}\right)^{-1}\left(m\right)$.
Moreover, a point $m\in\mathcal{N}_{0}$ is a transverse infinitesimal
intersection for $\mathcal{N}_{\varepsilon}^{+}$ and $\mathcal{N}_{\varepsilon}^{-}$
iff it is so for $\mathcal{N}_{\varepsilon}$ and $\mathcal{N}_{0}$.
Indeed, the flow generating $\mathcal{N}_{\varepsilon}$ is $\left(\phi_{X_{\varepsilon}^{-}}^{\varepsilon}\right)^{-1}\circ\phi_{X_{\varepsilon}^{+}}^{\varepsilon}$
whose vector field, denoted by $X_{\varepsilon}$, is equal to $\left(\phi_{X_{\varepsilon}^{-}}^{\varepsilon}\right)_{*}^{-1}\left(X_{\varepsilon}^{+}-X_{\varepsilon}^{-}\right)$.
For $\varepsilon=0$, one has simply $X_{0}=X_{0}^{+}-X_{0}^{-}$,
which proves that the operators $\left[.,X_{0}^{+}-X_{0}^{-}\right]$
and $\left[.,X_{0}-0\right]$ coincide.
\item Then, consider the fibrewise dilation by a constant number $\frac{1}{\varepsilon}$,
which is a diffeomorphism of $E$ and leaves the zero-section $\mathcal{N}_{0}$
invariant. This means that $\tilde{\alpha}_{\varepsilon}=\frac{\alpha_{\varepsilon}}{\varepsilon}$
is still a section, and it is smooth with respect to $\varepsilon$
even at $\varepsilon=0$ since $\alpha_{0}=0$. Namely, one has $\tilde{\alpha}_{0}=\left.\frac{d\alpha_{\varepsilon}}{d\varepsilon}\right|_{\varepsilon=0}$
which is nothing but $X_{0}$, provided the fibres are identified
with their tangent space. We denote by $\tilde{\mathcal{N}}_{\varepsilon}$
the graph of the section $\tilde{\alpha}_{\varepsilon}$, which is
thus a smooth family of manifolds. Since the dilatation is a diffeomorphism
for all $\varepsilon\neq0$, then $\mathcal{N}_{\varepsilon}$ and
$\mathcal{N}_{0}$ intersect transversally for all $\varepsilon\neq0$
iff $\tilde{\mathcal{N}}_{\varepsilon}$ and $\mathcal{N}_{0}$ do.
Now, we know from the general transversality theory that if $\tilde{\mathcal{N}}_{0}$
and $\mathcal{N}_{0}$ intersect transversally at some point $m$,
then for small enough $\varepsilon$ the intersection of $\tilde{\mathcal{N}}_{\varepsilon}$
and $\mathcal{N}_{0}$ near $m$ is a smooth manifold depending smoothly
on $\varepsilon$.
\item Finally, we show that if $m\in\mathcal{N}_{0}$ is a {}``transverse
infinitesimal intersection'' of $\mathcal{N}_{\varepsilon}$ and
$\mathcal{N}_{0}$ in the sense of Definition \ref{def_transverse},
then it is actually a transverse intersection (in the usual sense)
of $\tilde{\mathcal{N}}_{0}$ and $\mathcal{N}_{0}$. This can easily
be deduced from the following formula\begin{equation}
\left(\tilde{\alpha}_{0}\right)_{*}\left(Y\right)=\iota_{*}Y+D_{m,X_{0}}Y\label{eq_1}\end{equation}
which holds for each $Y\in T_{m}\mathcal{N}_{0}$. To show this formula,
we first use $\tilde{\alpha}_{0}=\phi_{X_{0}}^{1}\circ\iota$ and
thus $\left(\tilde{\alpha}_{0}\right)_{*}\left(Y\right)=\left(\phi_{X_{0}}^{1}\right)_{*}\iota_{*}Y$.
Let $\tilde{Y}\in\Gamma\left(TE\right)$ be any extension to $E$
of $\iota_{*}Y$, i.e., a vector field on $E$ satisfying $\iota_{*}Y_{m}=\tilde{Y}_{m}$.
We have \[
\left(\phi_{X_{0}}^{1}\right)_{*}\tilde{Y}=\tilde{Y}+\int_{0}^{1}\frac{d}{dt}\left(\left(\phi_{X_{0}}^{t}\right)_{*}\tilde{Y}\right)\, dt.\]
By definition of the Lie bracket, we obtain \[
\left(\phi_{X_{0}}^{1}\right)_{*}\tilde{Y}=\tilde{Y}+\int_{0}^{1}\left(\phi_{X_{0}}^{t}\right)_{*}\left[\tilde{Y},X_{0}\right]\, dt.\]
Let's choose $\tilde{Y}$ to be a lift of a vector field on the base
$\mathcal{N}_{0}$. Since $X_{0}$ is vertical, it follows that the
Lie bracket $\left[\tilde{Y},X_{0}\right]$ is also vertical, as well
as $\left(\phi_{X_{0}}^{t}\right)_{*}\left[\tilde{Y},X_{0}\right]$.
If $m\in\mathcal{N}_{0}$ is an infinitesimal intersection of $\mathcal{N}_{\varepsilon}$
and $\mathcal{N}_{0}$, i.e., a point where $\tilde{\alpha}_{0}$
vanishes, then the vector field $X_{0}$ vanishes everywhere on the
fibre above $m$ and the flow $\phi_{X_{0}}^{t}$ restricted to this
fibre $\mathcal{M}_{q}$ is the identity for all $t$. Thus, at such
a point $m$, one has $\int_{0}^{1}\left(\phi_{X_{0}}^{t}\right)_{*}\left[\tilde{Y},X_{0}\right]\, dt=\left[\tilde{Y},X_{0}\right]_{m}=D_{q,X_{0}}\left(Y\right)$
which proves the formula (\ref{eq_1}).
\end{itemize}
\end{proof}

\subsection{Intersections with constraints\label{sec_constraint}}

Suppose now that the two families $\mathcal{N}_{\varepsilon}^{\pm}$
are constrained to an intermediate compact submanifold $\mathcal{P}_{\varepsilon}$
for all $\varepsilon$, i.e., $\mathcal{N}_{\varepsilon}^{\pm}\subset\mathcal{P}_{\varepsilon}\subset\mathcal{M}$,
where $\mathcal{P}_{\varepsilon}$ is a smooth family of submanifolds
of codimension at least $1$. The submanifolds $\mathcal{N}_{\varepsilon}^{\pm}$
are thus in no way transverse in $\mathcal{M}$ but they may be so
in $\mathcal{P}_{\varepsilon}$ if an appropriate infinitesimal transversality
condition is satisfied, as we prove in Theorem \ref{theo_contraint_transverse_intersect}.
But first of all, we prove the following.

\begin{lem}
\label{lem_constraint_generating_flow}Let $\mathcal{N}_{\varepsilon}^{\pm}\subset\mathcal{P}_{\varepsilon}\subset\mathcal{M}$
be two families of constrained compact submanifolds. There exist generating
vector fields $X_{\varepsilon}^{\pm}$ which generate $\mathcal{P}_{\varepsilon}$
in the same time.
\end{lem}
\begin{proof}
First, let $\psi_{\varepsilon}$ be a flow generating the family $\mathcal{P}_{\varepsilon}$
and consider the families $\tilde{\mathcal{N}}_{\varepsilon}^{\pm}:=\psi_{\varepsilon}^{-1}\left(\mathcal{N}_{\varepsilon}^{\pm}\right)$.
These families satisfy $\tilde{\mathcal{N}}_{0}^{\pm}=\mathcal{N}_{0}$
and they are included in the fixed manifold $\mathcal{P}_{0}$ since
$\psi_{\varepsilon}\left(\mathcal{P}_{0}\right)=\mathcal{P}_{\varepsilon}$.
Therefore, there exist generating flows for $\tilde{\mathcal{N}}_{\varepsilon}^{\pm}$
inside $\mathcal{P}_{0}$, i.e., families of diffeomorphisms $\varphi_{\varepsilon}^{\pm}:\mathcal{P}_{0}\rightarrow\mathcal{P}_{0}$,
with vector fields $Y_{\varepsilon}^{\pm}$, such that $\varphi_{\varepsilon}^{\pm}\left(\mathcal{N}_{0}\right)=\tilde{\mathcal{N}}_{\varepsilon}^{\pm}$
and $Y_{0}^{+}-Y_{0}^{-}$ is not tangent to $\mathcal{N}_{0}$. Extending
these flows to families of diffeomorphisms on $\mathcal{M}$, we obtain
generating flows $\chi_{\varepsilon}^{\pm}$ of $\tilde{\mathcal{N}}_{\varepsilon}^{\pm}$
on $\mathcal{M}$ with the property that $\chi_{\varepsilon}^{\pm}\left(\mathcal{P}_{0}\right)=\mathcal{P}_{0}$,
and with vector fields $Z_{\varepsilon}^{\pm}$ satisfying $Z_{0}^{+}-Z_{0}^{-}$
not tangent to $\mathcal{N}_{0}$. This implies that the families
$\phi_{\varepsilon}^{\pm}:=\psi_{\varepsilon}\circ\chi_{\varepsilon}^{\pm}$
generate $\mathcal{N}_{\varepsilon}^{\pm}$ and satisfy $\phi_{\varepsilon}^{\pm}\left(\mathcal{P}_{0}\right)=\mathcal{P}_{\varepsilon}$. 
\end{proof}
\begin{defn}
\label{def_infinit_intersect_constraint}Let $\mathcal{N}_{\varepsilon}^{\pm}\subset\mathcal{P}_{\varepsilon}\subset\mathcal{M}$
be two families of constrained compact submanifolds. An infinitesimal
intersection $m\in\mathcal{N}_{0}$ is called \textbf{transverse in
the constraint} whenever $\textrm{img}D_{m,X_{0}^{\pm}}$ is transverse
to $T_{m}\mathcal{N}_{0}$ in $T_{m}\mathcal{P}_{0}$, i.e.,\[
\textrm{img}D_{m,X_{0}^{\pm}}\oplus T_{m}\mathcal{N}_{0}=T_{m}\mathcal{P}_{0},\]
with $X_{\varepsilon}^{\pm}$ any generating vector fields.
\end{defn}
To check that this notion is well-defined, one has to verify two facts.
First, the image $\textrm{img}D_{m,X_{0}^{\pm}}$ is in $T_{m}\mathcal{P}_{0}$.
Indeed, for any $Y\in T_{m}\mathcal{N}_{0}$ we can choose an extension
$\tilde{Y}$ which is tangent to both $\mathcal{N}_{0}$ and $\mathcal{P}_{0}$,
hence $D_{m,\tilde{X}_{0}^{\pm}}\left(Y\right)=\left[\tilde{Y},X_{0}^{+}-X_{0}^{-}\right]_{m}$
lies in $\mathcal{P}_{0}$ since $\tilde{Y}$ and $X_{0}^{+}-X_{0}^{-}$
do. Second, this notion of transversality is independent of the choice
of the generating vector fields $X_{\varepsilon}^{\pm}$, as one can
check easily following the proof of Lemma \ref{lem_transverse_bien_def}.

As before, we have an equivalent criterion for transverse infinitesimal
intersections, in terms of $\dim\left(\textrm{ker}D_{m,X_{0}^{\pm}}\right)$.
With the help of Lemma \ref{lem_constraint_generating_flow}, Lemma
\ref{lem_transverse_intersect_dim_noyau} transposes straightforwardly
to the context with constraint, as follows.

\begin{lem}
\label{lem_constraint_transverse_intersect_dim_noyau}Let $\mathcal{N}_{\varepsilon}^{\pm}\subset\mathcal{P}_{\varepsilon}\subset\mathcal{M}$
be two families of constrained compact submanifolds. There exist generating
vector fields $X_{\varepsilon}^{\pm}$ such that for any infinitesimal
intersection $m\in\mathcal{N}_{0}$, the space $\textrm{img}\left(D_{m,X_{0}^{\pm}}\right)$
does not intersect $T_{m}\mathcal{N}_{0}$. For such $X_{\varepsilon}^{\pm}$,
$m$ is transverse in the constraint iff\[
\dim\left(\textrm{ker}D_{m,X_{0}^{\pm}}\right)=2\dim\mathcal{N}_{0}-\dim\mathcal{P}_{0}.\]

\end{lem}
\begin{thm}
\label{theo_contraint_transverse_intersect}Let $\mathcal{N}_{\varepsilon}^{\pm}\subset\mathcal{P}_{\varepsilon}\subset\mathcal{M}$
be two families of constrained compact submanifolds. If $m\in\mathcal{N}_{0}$
is a transverse infinitesimal intersection in the constraint, then
in a neighbourhood of $m$ there is a smooth family of submanifolds
$\Lambda_{\varepsilon}$ with $\Lambda_{0}\in\mathcal{N}_{0}$ and
$\Lambda_{\varepsilon}=\mathcal{N}_{\varepsilon}^{+}\cap\mathcal{N}_{\varepsilon}^{-}$
for small enough $\varepsilon\neq0$.
\end{thm}
\begin{proof}
Let's choose a flow $\psi_{\varepsilon}$ generating $\mathcal{P}_{\varepsilon}$
and denote by $Z_{\varepsilon}$ its associated vector field. First,
one proves that a point $m\in\mathcal{N}_{0}$ is a \textbf{}transverse
infinitesimal intersection of $\mathcal{N}_{\varepsilon}^{+}$ and
$\mathcal{N}_{\varepsilon}^{-}$ in the constraint $\mathcal{P}_{\varepsilon}$
iff it is a transverse infinitesimal intersection of $\psi_{\varepsilon}^{-1}\left(\mathcal{N}_{\varepsilon}^{+}\right)$
and $\psi_{\varepsilon}^{-1}\left(\mathcal{N}_{\varepsilon}^{-}\right)$
in the constraint $\mathcal{P}_{0}$, where $\mathcal{P}_{0}$ is
understood here as the constant family $\mathcal{P}_{\varepsilon}=\mathcal{P}_{0}$.
Indeed, let's define $\tilde{\mathcal{N}}_{\varepsilon}^{\pm}:=\psi_{\varepsilon}^{-1}\left(\mathcal{N}_{\varepsilon}^{\pm}\right)$.
Suppose that $m$ is a transverse infinitesimal intersection of $\mathcal{N}_{\varepsilon}^{+}$
and $\mathcal{N}_{\varepsilon}^{-}$ in the constraint $\mathcal{P}_{\varepsilon}$,
i.e., $\textrm{img}D_{m,X_{0}^{\pm}}\oplus T_{m}\mathcal{N}_{0}=T_{m}\mathcal{P}_{0}$,
where $X_{\varepsilon}^{\pm}$ generates $\mathcal{N}_{\varepsilon}^{\pm}$.
The family $\psi_{\varepsilon}^{-1}\left(\mathcal{N}_{\varepsilon}^{\pm}\right)$
is generated by the flow $\psi_{\varepsilon}^{-1}\circ\phi_{X_{\varepsilon}^{\pm}}^{\varepsilon}$
whose associated vector field $\tilde{X}_{\varepsilon}^{\pm}$ equals
to $\left(\psi_{\varepsilon}^{-1}\right)_{*}\left(-Z_{\varepsilon}+X_{\varepsilon}^{\pm}\right)$.
For $\varepsilon=0$, this reduces $\tilde{X}_{0}^{\pm}=-Z_{0}+X_{0}^{\pm}$.
Consequently, we have $\tilde{X}_{0}^{+}-\tilde{X}_{0}^{-}=X_{0}^{+}-X_{0}^{-}$
and thus $D_{m,\tilde{X}_{0}^{\pm}}=D_{m,X_{0}^{\pm}}$. Since $\tilde{\mathcal{N}}_{0}=\mathcal{N}_{0}$,
we have shown that $\textrm{img}D_{m,X_{0}^{\pm}}\oplus T_{m}\mathcal{N}_{0}=T_{m}\mathcal{P}_{0}$
is equivalent to $\textrm{img}D_{m,\tilde{X}_{0}^{\pm}}\oplus T_{m}\mathcal{N}_{0}=T_{m}\mathcal{P}_{0}$.

Now, since the families $\tilde{\mathcal{N}}_{\varepsilon}^{\pm}$
lie in the fixed submanifold $\mathcal{P}_{0}$, we can apply Theorem
\ref{theo_2manifold_transverse_intersect} which insures that near
$m$ there is a smooth family of submanifolds $\tilde{\Lambda}_{\varepsilon}$
with $\tilde{\Lambda}_{0}\subset\mathcal{N}_{0}$ and $\tilde{\Lambda}_{\varepsilon}=\tilde{\mathcal{N}}_{\varepsilon}^{+}\cap\tilde{\mathcal{N}}_{\varepsilon}^{-}$
for small enough $\varepsilon\neq0$. Applying then the family of
diffeomorphisms $\psi_{\varepsilon}$, we obtain the claimed result
for the intersections of the families $\mathcal{N}_{\varepsilon}^{\pm}$.
\end{proof}

\subsection{Lagrangian intersections\label{sec_lagrangian_intersection}}

Let us suppose now that $\mathcal{M}$ is endowed with a symplectic
structure $\omega$ and that the families of submanifolds $\mathcal{N}_{\varepsilon}^{\pm}$
are Lagrangian for all $\varepsilon$.

\subsubsection{Mel'nikov $1$-form for pairs of Lagrangian submanifolds}

\begin{defn}
\label{def_melnikov_1form}Let $\mathcal{N}_{\varepsilon}^{\pm}\subset\mathcal{M}$
be two families of compact Lagrangian submanifolds. The \textbf{Mel'nikov
$1$-form} $\beta\in\Omega^{1}\left(\mathcal{N}_{0}\right)$ is defined
by \[
\beta:=\iota^{*}\left(\left(X_{0}^{+}-X_{0}^{-}\right)\lrcorner\omega\right),\]
 where $\iota:\mathcal{N}_{0}\hookrightarrow\mathcal{M}$ is the inclusion
map and $X_{\varepsilon}^{\pm}$ are any generating vector fields.
\end{defn}
\begin{lem}
\label{lem_melnikov_is_closed}The Mel'nikov $1$-form is well-defined,
i.e., independent of the choice of $X_{\varepsilon}^{\pm}$, and it
is a closed form, $d\beta=0$.
\end{lem}
\begin{proof}
First, if $\tilde{X}_{\varepsilon}^{\pm}$ is a second pair of generating
vector fields, we know from Lemma \ref{lem_generating_flow_non_unique}
that both differences $X_{0}^{+}-\tilde{X}_{0}^{+}$ and $X_{0}^{-}-\tilde{X}_{0}^{-}$
are tangent to $\mathcal{N}_{0}$, and therefore so is the vector
field $Z=X_{0}^{+}-X_{0}^{-}-\left(\tilde{X}_{0}^{+}-\tilde{X}_{0}^{-}\right)$.
If we denote by $\beta$ (resp. $\tilde{\beta}$ ) the Mel'nikov $1$-form
defined with $X_{\varepsilon}^{\pm}$ (resp. $\tilde{X}_{\varepsilon}^{\pm}$),
we have the relation $\beta=\tilde{\beta}+\iota^{*}\left(Z\lrcorner\omega\right)$.
The second term vanishes since $Z$ is tangent to $\mathcal{N}_{0}$
which is Lagrangian and therefore $\beta=\tilde{\beta}$.

Second, for each $\varepsilon$ the pull-back $\iota^{*}\left(\phi_{X_{\varepsilon}^{\pm}}^{\varepsilon}\right)^{*}\omega$
vanishes on $\mathcal{N}_{0}$ since $\phi_{X_{\varepsilon}^{\pm}}^{\varepsilon}\circ\iota\left(\mathcal{N}_{0}\right)=\mathcal{N}_{\varepsilon}^{\pm}$
and the manifolds $\mathcal{N}_{\varepsilon}^{\pm}$ are Lagrangian.
Taking the derivative with respect to $\varepsilon$ and using Cartan's
formula together with $d\omega=0$, one obtains $\iota^{*}\left(\phi_{X_{\varepsilon}^{\pm}}^{\varepsilon}\right)^{*}d\left(X_{\varepsilon}^{\pm}\lrcorner\omega\right)=0$,
i.e., $d\left(\iota^{*}\left(\phi_{X_{\varepsilon}^{\pm}}^{\varepsilon}\right)^{*}\left(X_{\varepsilon}^{\pm}\lrcorner\omega\right)\right)=0$.
Then, for $\varepsilon=0$ one has $d\left(\iota^{*}\left(X_{0}^{\pm}\lrcorner\omega\right)\right)=0$
and the difference between the term with $X_{\varepsilon}^{+}$ and
the one with $X_{\varepsilon}^{-}$ gives exactly $d\beta=0$.
\end{proof}
In this symplectic context, one can conveniently reformulate the infinitesimal
transversality condition in terms of $\beta$ instead of $X_{0}^{+}-X_{0}^{-}$. 

\begin{lem}
\label{lem_infinit_intersect_melnikov}Let $\mathcal{N}_{\varepsilon}^{\pm}\subset\mathcal{M}$
be two families of compact Lagrangian submanifolds and $\beta$ the
Mel'nikov $1$-form. A point $m\in\mathcal{N}_{0}$ is an infinitesimal
intersection iff $\beta$ vanishes at $m$.
\end{lem}
\begin{proof}
The {}``only if'' part of the assumption is obvious. In order to
prove the {}``if'' part, let us assume that $\beta=0$ at the point
$m$. By definition, this means that $\omega\left(X_{0}^{+}-X_{0}^{-},\iota_{*}Z\right)=0$
for all $Z\in T_{m}\mathcal{N}_{0}$. This implies that $X_{0}^{+}-X_{0}^{-}$
is in the $\omega$-orthogonal of $T_{m}\mathcal{N}_{0}$ which is
$T_{m}\mathcal{N}_{0}$ itself, since $\mathcal{N}_{0}$ is Lagrangian.
But, $X_{0}^{+}-X_{0}^{-}$ is by assumption (Lemma \ref{lem_generating_flow_exist})
never tangent to $\mathcal{N}_{0}$. Therefore $X_{0}^{+}-X_{0}^{-}=0$
at $m$.
\end{proof}
\begin{lem}
\label{lem_hessian_symmetric}Let $\mathcal{N}_{\varepsilon}^{\pm}\subset\mathcal{M}$
be two families of compact Lagrangian submanifolds and $\beta$ the
Mel'nikov $1$-form. If $m\in\mathcal{N}_{0}$ is an infinitesimal
intersection, then the derivative $\nabla\beta:T_{m}\mathcal{N}_{0}\times T_{m}\mathcal{N}_{0}\rightarrow\mathbb{R}$
defined by\[
\left(\nabla\beta\right)\left(Y,Z\right)=Y\left(\beta\left(\tilde{Z}\right)\right),\]
 with $\tilde{Z}\in\Gamma\left(T\mathcal{N}_{0}\right)$ any extension
of $Z$, is a well-defined symmetric bilinear form.
\end{lem}
\begin{proof}
Indeed, by definition of the Lie derivative, one has $Y\left(\beta\left(\tilde{Z}\right)\right)=\mathcal{L}_{\tilde{Y}}\left(\beta\left(\tilde{Z}\right)\right)$,
where $\tilde{Y}$ is any extension on $\mathcal{N}_{0}$ of $Y$.
Then, the Leibniz rule gives $Y\left(\beta\left(\tilde{Z}\right)\right)=\tilde{Z}\lrcorner\mathcal{L}_{\tilde{Y}}\beta+\left(\mathcal{L}_{\tilde{Y}}\tilde{Z}\right)\lrcorner\beta$.
The second term vanishes at the point $m$ since $\beta$ does. Then,
applying Cartan's formula to the first term, we obtain \[
Y\left(\beta\left(\tilde{Z}\right)\right)=\tilde{Z}\lrcorner\left(\tilde{Y}\lrcorner d\beta+d\left(\beta\left(\tilde{Y}\right)\right)\right).\]
 The first term vanishes since $\beta$ is closed. We thus have $Y\left(\beta\left(\tilde{Z}\right)\right)=Z\left(\beta\left(\tilde{Y}\right)\right)$
which is independent of the choice of the extension $\tilde{Z}$.
\end{proof}
We remark that the use of the symbol $\nabla$ is well-justified since
the derivative $\left(\nabla\beta\right)\left(Y,Z\right)$ is easily
shown to be equal to $\left(\nabla^{'}\beta\right)\left(\tilde{Y},\tilde{Z}\right)_{m}$,
where $\nabla^{'}$ is any covariant derivative and $\tilde{Y},\tilde{Z}$
are any extensions to $\mathcal{N}_{0}$ of $Y,Z$. The derivative
$\nabla\beta$ is related to $D_{m,X_{0}^{\pm}}$ as follows.

\begin{lem}
\label{lem_relation_operator_D_melnikov}Let $\mathcal{N}_{\varepsilon}^{\pm}\subset\mathcal{M}$
be two families of compact Lagrangian submanifolds, $X_{\varepsilon}^{\pm}$
generating vector fields and $\beta$ the Mel'nikov $1$-form. For
any infinitesimal intersection $m\in\mathcal{N}_{0}$, we have the
following relation\[
\left(\nabla\beta\right)\left(Y,Z\right)=\omega\left(D_{m,X_{0}^{\pm}}\left(Y\right),i_{*}Z\right),\]
 for all $Y,Z\in T_{m}\mathcal{N}_{0}$.
\end{lem}
\begin{proof}
By definition, one has $\left(\nabla\beta\right)\left(Y,Z\right)=\mathcal{L}_{\tilde{Y}}\left(\beta\left(\tilde{Z}\right)\right)$,
with $\tilde{Y}\in\Gamma\left(T\mathcal{N}_{0}\right)$ (resp. $\tilde{Z}\in\Gamma\left(T\mathcal{N}_{0}\right)$)
any extension of $Y$ (resp. $Z$). The Leibniz rule gives $Y\left(\beta\left(\tilde{Z}\right)\right)=Z\lrcorner\mathcal{L}_{\tilde{Y}}\beta+\left(\mathcal{L}_{\tilde{Y}}\tilde{Z}\right)\lrcorner\beta$.
The second term vanishes at the point $m$ since $\beta$ does and
introducing the definition of $\beta$ in the first term gives $Y\left(\beta\left(\tilde{Z}\right)\right)=Z\lrcorner\mathcal{L}_{\tilde{Y}}\left(\iota^{*}\left(\left(X_{0}^{+}-X_{0}^{-}\right)\lrcorner\omega\right)\right)$.
If we choose any extension $Y^{'}$ on $\mathcal{M}$ of $\iota_{*}\tilde{Y}$,
we have $Y\left(\beta\left(\tilde{Z}\right)\right)=Z\lrcorner\iota^{*}\left(\mathcal{L}_{Y^{'}}\left(\left(X_{0}^{+}-X_{0}^{-}\right)\lrcorner\omega\right)\right)$.
Using once again the Leibniz rule provides \[
Y\left(\beta\left(\tilde{Z}\right)\right)=Z\lrcorner\iota^{*}\left(\left[Y^{'},X_{0}^{+}-X_{0}^{-}\right]\lrcorner\omega+\left(X_{0}^{+}-X_{0}^{-}\right)\lrcorner\mathcal{L}_{Y^{'}}\omega\right).\]
 The second term vanishes at $m$ since $X_{0}^{+}-X_{0}^{-}$ does
and the first term is precisely $Z\lrcorner\iota^{*}\left(D_{m,X_{0}^{\pm}}\left(Y\right)\lrcorner\omega\right)$,
i.e., $\omega\left(D_{m,X_{0}^{\pm}}\left(Y\right),\iota_{*}Z\right)$.
\end{proof}
This equality has the following corollary.

\begin{lem}
\label{lem_noyau_D_egal_noyau_hessian}Let $\mathcal{N}_{\varepsilon}^{\pm}\subset\mathcal{M}$
be two families of compact Lagrangian submanifolds and $\beta$ the
Mel'nikov $1$-form. There exist generating vector fields $X_{\varepsilon}^{\pm}$
such that for any infinitesimal intersection $m\in\mathcal{N}_{0}$,
the space $\textrm{img}\left(D_{m,X_{0}^{\pm}}\right)$ does not intersect
$T_{m}\mathcal{N}_{0}$. For such $X_{\varepsilon}^{\pm}$, we have
the relation\[
\textrm{ker}D_{m,X_{0}^{\pm}}=\textrm{ker}\nabla\beta.\]

\end{lem}
\begin{proof}
Indeed, thanks to the relation given in Lemma \ref{lem_relation_operator_D_melnikov},
we see that $\textrm{ker}D_{m,X_{0}^{\pm}}\subset\textrm{ker}\nabla\beta$.
The converse inclusion $\textrm{ker}D_{m,X_{0}^{\pm}}\supset\textrm{ker}\nabla\beta$
is proved as follows. First, the existence of generating vector fields
$X_{\varepsilon}^{\pm}$ with the announced property is proved in
\ref{lem_transverse_intersect_dim_noyau}. Therefore, if $\left(\nabla\beta\right)\left(Y,Z\right)=0$
for all $Z\in T_{m}\mathcal{N}_{0}$ then $D_{m,X_{0}^{\pm}}\left(Y\right)$
must lie in the $\omega$-orthogonal of $T_{m}\mathcal{N}_{0}$, which
is $T_{m}\mathcal{N}_{0}$ itself. But this is a contradiction and
therefore $D_{m,X_{0}^{\pm}}\left(Y\right)$ must vanish.
\end{proof}

\subsubsection{Constrained intersections for Lagrangian submanifolds\label{sec_lagrangian_with_constraints}}

We suppose now that our Lagrangian submanifolds $\mathcal{N}_{\varepsilon}^{\pm}$
are constrained to an intermediate submanifold $\mathcal{P}_{\varepsilon}$
for all $\varepsilon$, as described on Section \ref{sec_constraint}.
Thanks to Lemma \ref{lem_noyau_D_egal_noyau_hessian}, the criterion
given in Lemma \ref{lem_constraint_transverse_intersect_dim_noyau}
transposes straightforwardly to the Lagrangian case, as follows.

\begin{lem}
\label{lem_constraint_lag_transverse_intersect_dim_noyau}Let $\mathcal{N}_{\varepsilon}^{\pm}\subset\mathcal{P}_{\varepsilon}\subset\mathcal{M}$
be two families of constrained compact Lagrangian submanifolds and
$\beta$ be the Mel'nikov $1$-form. Then, an infinitesimal intersection
$m$ is transverse in the constraint iff\[
\dim\left(\textrm{ker}\nabla\beta\right)=\textrm{codim}\mathcal{P}_{0}.\]

\end{lem}
We can actually say more than this. Indeed, since $\mathcal{P}_{0}$
contains the Lagrangian manifold $\mathcal{N}_{0}$, it must be coisotropic
and the associated isotropic foliation $\left(T_{m}\mathcal{P}_{0}\right)^{\perp}$satisfies
$\left(T_{m}\mathcal{P}_{0}\right)^{\perp}\subset T_{m}\mathcal{N}_{0}$
for all $m\in\mathcal{N}_{0}$. Moreover, the dimension of the isotropic
foliation is exactly equal to $\textrm{codim}\mathcal{P}_{0}$. This
allows to show Proposition \ref{prop_constrained_lagrangian_transverse_condition}
which will be easily deduced from the following lemma.

\begin{lem}
\label{lem_beta_evaluee_sur_XH0}Let $\mathcal{N}_{\varepsilon}^{\pm}\subset\mathcal{M}$
be two families of compact Lagrangian submanifolds and $\beta$ the
Mel'nikov $1$-form. If $F_{\varepsilon}\in C^{\infty}\left(\mathcal{M}\right)$
is a family of smooth functions constant on $\mathcal{N}_{\varepsilon}^{+}$
and $\mathcal{N}_{\varepsilon}^{-}$ for all $\varepsilon$, then
the Hamiltonian vector field $X_{F_{0}}$ is tangent to $\mathcal{N}_{0}$
and satisfies \[
\beta\lrcorner X_{F_{0}}=0\]
everywhere on $\mathcal{N}_{0}$.
\end{lem}
\begin{proof}
Let $X_{\varepsilon}^{\pm}$ be vector fields generating the families
$\mathcal{N}_{\varepsilon}^{\pm}$. The Mel'nikov $1$-form is related
to them by $\beta=\iota^{*}\left(\left(X_{0}^{+}-X_{0}^{-}\right)\lrcorner\omega\right)$.
By hypothesis, there exists a family of real numbers $c_{\varepsilon}$
such that $F_{\varepsilon}\circ\phi_{X_{\varepsilon}^{\pm}}^{\varepsilon}\circ\iota=c_{\varepsilon}$
for all $\varepsilon$, where $\iota:\mathcal{N}_{0}\hookrightarrow\mathcal{M}$
is the inclusion map. Taking the derivative with respect to $\varepsilon$,
one obtains\[
\left(\frac{F_{\varepsilon}}{d\varepsilon}+X_{\varepsilon}^{\pm}\left(F_{\varepsilon}\right)\right)\circ\phi_{X_{\varepsilon}^{\pm}}^{\varepsilon}\circ\iota=\frac{dc_{\varepsilon}}{d\varepsilon}.\]
Denoting by a dot the derivatives with respect to $\varepsilon$,
one has \[
\left(\dot{F}_{0}+X_{0}^{\pm}\left(F_{0}\right)\right)\circ\iota=\dot{c}_{0},\]
since $\left(\phi_{X_{\varepsilon}^{\pm}}^{\varepsilon}\right)_{\varepsilon=0}=\mathbb{I}$.
The difference between the term with $X_{0}^{+}$ and the one with
$X_{0}^{-}$ gives simply $\left(X_{0}^{+}-X_{0}^{-}\right)\left(F_{0}\right)\circ\iota=0$.
Now, by definition of the Hamiltonian vector field $X_{F_{0}}$, the
function $\left(X_{0}^{+}-X_{0}^{-}\right)\left(F_{0}\right)$ is
equal to $\omega\left(X_{0}^{+}-X_{0}^{-},X_{F_{0}}\right)$. Moreover,
$X_{F_{0}}$ is tangent to $\mathcal{N}_{0}$ because at each $m\in\mathcal{N}_{0}$,
the Lagrangian space $T_{m}\mathcal{N}_{0}$ is included in $\ker\left(dF_{0}\right)_{m}$.
This implies that $\omega\left(X_{0}^{+}-X_{0}^{-},X_{F_{0}}\right)$
is simply $\beta\left(X_{F_{0}}\right)$ and the result follows.
\end{proof}
\begin{prop}
\label{prop_constrained_lagrangian_transverse_condition}Let $\mathcal{N}_{\varepsilon}^{\pm}\subset\mathcal{P}_{\varepsilon}\subset\mathcal{M}$
be two families of constrained compact Lagrangian submanifolds and
$\beta$ the Mel'nikov $1$-form. Then an infinitesimal intersection
$m\in\mathcal{N}_{0}$ is transverse in the constraint iff \[
\textrm{ker}\nabla\beta=\left(T_{m}\mathcal{P}_{0}\right)^{\perp}.\]

\end{prop}
\begin{proof}
First, there exist $p$ smooth families of linearly independent functions
$F_{\varepsilon}^{\left(1\right)},...,F_{\varepsilon}^{\left(p\right)}\in C^{\infty}\left(\mathcal{M}\right)$,
where $p=\textrm{codim}\mathcal{P}_{\varepsilon}$, such that in a
neighbourhood of $m$ the manifold $\mathcal{P}_{\varepsilon}$ is
given by the common level set $\mathcal{P}_{\varepsilon}=\left\{ m\mid F_{\varepsilon}^{\left(1\right)}\left(m\right)=c_{\varepsilon}^{\left(1\right)},...,F_{\varepsilon}^{\left(p\right)}\left(m\right)=c_{\varepsilon}^{\left(p\right)}\right\} $,
where $c_{\varepsilon}^{\left(j\right)}$ are families of real numbers.
Applying the preceding lemma, we obtain that $X_{F_{0}^{\left(j\right)}}$
is tangent to $\mathcal{N}_{0}$ and satisfies $\beta\lrcorner X_{F_{0}^{\left(j\right)}}=0$
everywhere on $\mathcal{N}_{0}$, for each $j=1..p$. Now, at each
$m\in\mathcal{N}_{0}$ the vectors $X_{F_{0}^{\left(j\right)}}$ form
a basis of $\left(T_{m}\mathcal{P}_{0}\right)^{\perp}$. This implies
that $\left(T_{m}\mathcal{P}_{0}\right)^{\perp}\subset\textrm{ker}\beta$
everywhere on $\mathcal{N}_{0}$. Therefore, if $m$ is an infinitesimal
intersection, then for each $Z\in\left(T_{m}\mathcal{P}_{0}\right)^{\perp}\subset T_{m}\mathcal{N}_{0}$,
one has $\left(\nabla\beta\right)\left(Y,Z\right)=Y\left(\beta\left(\tilde{Z}\right)\right)=0$
since we can choose the extension $\tilde{Z}$ to be everywhere in
$\left(T_{m}\mathcal{P}_{0}\right)^{\perp}$. For such a $\tilde{Z}$,
one has $\beta\left(\tilde{Z}\right)$ everywhere and therefore $\left(\nabla\beta\right)\left(Y,Z\right)=0$
for all $Y$. This shows that $\left(T_{m}\mathcal{P}_{0}\right)^{\perp}\subset\textrm{ker}\nabla\beta$.
This inclusion together with the transversality condition $\dim\left(\textrm{ker}\nabla\beta\right)=\dim\left(T_{m}\mathcal{P}_{0}\right)^{\perp}$
of Lemma \ref{lem_constraint_transverse_intersect_dim_noyau} proves
the result.
\end{proof}

\section{The Mel'nikov 1-form\label{sec_melnikov}}

In the previous section, we developed tools to deal with pairs $\mathcal{N}_{\varepsilon}^{\pm}$
of families of Lagrangian submanifolds, with the same limit $\mathcal{N}_{0}:=\mathcal{N}_{0}^{+}=\mathcal{N}_{0}^{-}$
and constrained for all $\varepsilon$ to a submanifold $\mathcal{P}_{\varepsilon}$.
We will now use these tools to deal with the situation where $\mathcal{N}_{\varepsilon}^{\pm}$
are respectively the stable and unstable manifolds of transversally
hyperbolic periodic orbits of a given Hamiltonian on $\mathcal{M}$.
The Mel'nikov $1$-form introduced in Definition \ref{def_melnikov_1form}
allows us to detect the presence of intersections of $\mathcal{N}_{\varepsilon}^{-}$
and $\mathcal{N}_{\varepsilon}^{+}$, i.e., heteroclinic orbits between
the two periodic orbits. After setting precisely the heteroclinic
and homoclinic situation we will deal with, we show that the Mel'nikov
$1$-form admit an integral expression whenever the Hamiltonian is
completely integrable. This integral is unfortunately not convergent
in general and needs a prescription on the way we take the limit.
Nevertheless, we consider two cases in which this integral is convergent.
In particular, this encompasses the historical Mel'nikov setup (time-periodic
perturbation of time-independent systems) which is presented as a
conclusion of this paper.

\subsection{Heteroclinic and homoclinic motions }

\subsubsection{Stable and unstable manifolds of transversally hyperbolic orbits\label{sub_moebius}}

Suppose the dimension of $\mathcal{M}$ is at least $4$. Let $H\in C^{\infty}\left(\mathcal{M}\right)$
be a Hamiltonian and denote by $X_{H}$ its vector field and by $\phi^{t}$
its flow. We recall here some basic facts about stable and unstable
manifolds of transversally hyperbolic periodic orbit and refer e.g.
to \cite{abraham_marsden1} for more details.

\begin{defn}
\label{def_orbit_hyperbolic}A $\tau$-periodic orbit $\gamma$ of
$X_{H}$ is called \textbf{(transversally) non-degenerate} whenever
the eigenvalue $\lambda=1$ of the derivative map $\phi_{*{}}^{\tau}$
at some point $m\in\gamma$ has multiplicity $2$. If moreover the
other eigenvalues do not lie on the unit circle, $\gamma$ is called
\textbf{(transversally) hyperbolic.}
\end{defn}
Note that the eigenvalues of the map $\phi_{*}^{\tau}$ always come
in pairs $\left(\lambda,\lambda^{-1}\right)$ since $\phi^{\tau}$
is a symplectic map. On the other hand, at the point $m$ the vector
$X_{H}$ itself is obviously an eigenvector with eigenvalue $1$.

It is well-known that the nondegeneracy condition implies that such
a periodic orbit always arises within an orbit-cylinder $\Gamma$,
i.e., there is an embedding $\Gamma:S^{1}\times\left[a,b\right]\rightarrow\mathcal{M}$,
with $H\left(\gamma\right)\in\left[a,b\right]$, such that for each
$E\in\left[a,b\right]$, the circle $\gamma_{E}=\Gamma\left(S^{1}\times\left\{ E\right\} \right)$
is a closed orbit of $X_{H}$ and moreover $\Gamma$ is transversal
to the energy surfaces $\left\{ m;H\left(m\right)=E\right\} $.

Furthermore, the hyperbolicity of a periodic orbit $\gamma$ implies
the existence of the so-called \textbf{stable} and \textbf{unstable
manifolds}. The stable (resp. unstable) manifold is the set, denoted
by $\mathcal{N}^{+}$ (resp. $\mathcal{N}^{-}$) of points $m\in\mathcal{M}$
such that $\phi^{t}\left(m\right)$ tends to the limit cycle $\gamma$
when $t\rightarrow+\infty$ (resp. $t\rightarrow+\infty$). One is
usually obliged to distinguish between the \emph{local} and the \emph{global}
(un)stable manifolds. Indeed, the hyperbolicity condition implies
that in a neighbourhood of $\gamma$, there exist two embedded Lagrangian
submanifolds $\mathcal{N}_{loc}^{+}$ and $\mathcal{N}_{loc}^{-}$,
called the local stable and unstable manifolds, whose intersection
is exactly $\gamma$. The global stable and unstable manifolds are
then obtained from the local ones by applying the flow $\phi^{t}$
for all $t$, and in general they are injectively immersed in $\mathcal{M}$
in a very complicated way. 

\begin{center}\includegraphics{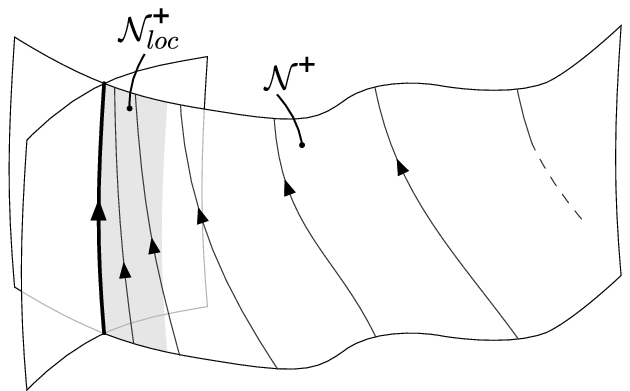}\end{center}

In the sequel, we will need to focus on a compact part of the stable
and unstable manifolds. For this purpose, we define the following.

\begin{defn}
\label{def_stable_manifold_T}For each $T>0$, we define the compact
manifold $\mathcal{N}_{T}^{\pm}:=\phi^{\mp T}\left(\overline{\mathcal{N}_{loc}^{\pm}}\right)$.
\end{defn}
These manifolds depend of course on the choice of the local manifolds
$\mathcal{N}_{loc}^{\pm}$ , but they satisfy $\mathcal{N}_{T}^{\pm}\subset\mathcal{N}_{T^{'}}^{\pm}$
for all $T<T^{'}$, and $\lim_{T\rightarrow+\infty}\mathcal{N}_{T}^{\pm}=\mathcal{N}^{\pm}$. 

We remark that in dimension $4$, the manifolds $\mathcal{N}_{loc}^{\pm}$
(and thus $\mathcal{N}_{T}^{\pm}$ as well) may be such that $\mathcal{N}_{loc}^{\pm}\setminus\gamma$
has two connected components, say $\mathcal{N}_{1}^{\pm}$ and $\mathcal{N}_{2}^{\pm}$,
as depicted on the left hand side below.

\begin{center}\begin{tabular}{ccc}
\includegraphics{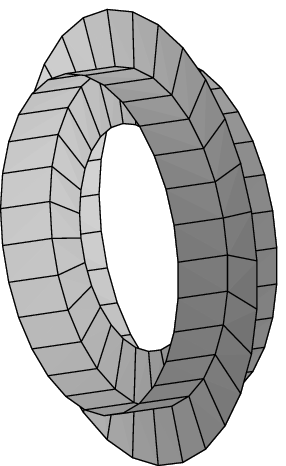}&
~~~~~&
\includegraphics{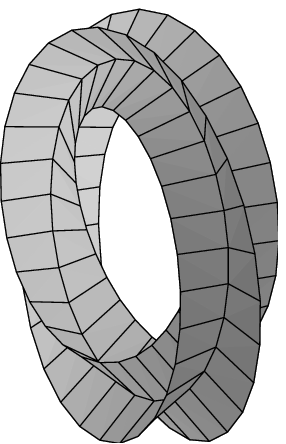}\tabularnewline
\end{tabular}\end{center}

In that case, $\mathcal{N}_{T}^{\pm}$ will rather denote $\phi^{\mp T}\left(\overline{\mathcal{N}_{j}^{\pm}}\right)$
with the choice of a connected component%
\footnote{See e.g. \cite{colin_vungoc} for a study of $4$-dimensional completely
integrable systems with transversally hyperbolic periodic orbits.%
} $j=1$ or $2$. As well, $\mathcal{N}^{\pm}$ will denote one component
of the (un)stable manifold rather than the full manifold. In higher
dimensions, this distinction is irrelevant since the manifolds $\mathcal{N}_{loc}^{\pm}\setminus\gamma$
are connected.

\subsubsection{Heteroclinic and homoclinic motions }

Let $H_{0}\in C^{\infty}\left(\mathcal{M}\right)$ be a Hamiltonian
which admits two hyperbolic periodic orbits $\gamma_{0}^{+}$ and
$\gamma_{0}^{-}$, and denote by $\phi^{t}$ its flow. As explained
in the previous section, the orbit $\gamma_{0}^{+}$ (resp. $\gamma_{0}^{-}$)
has a stable and an unstable manifold $\mathcal{N}^{\pm}\left(\gamma_{0}^{+}\right)$
(resp. $\mathcal{N}^{\pm}\left(\gamma_{0}^{-}\right)$). Let us focus
now on the two manifolds $\mathcal{N}^{+}\left(\gamma_{0}^{+}\right)$
and $\mathcal{N}^{-}\left(\gamma_{0}^{-}\right)$. Any point $m\in\mathcal{N}^{+}\left(\gamma_{0}^{+}\right)\cap\mathcal{N}^{-}\left(\gamma_{0}^{-}\right)$
is called a \textbf{heteroclinic point} and its orbit $t\rightarrow\phi^{t}\left(m\right)$
is a \textbf{heteroclinic orbit} between $\gamma_{0}^{-}$ and $\gamma_{0}^{+}$,
i.e., it tends to $\gamma_{0}^{-}$ (resp. $\gamma_{0}^{+}$) when
$t\rightarrow-\infty$ (resp. $t\rightarrow+\infty$). When the two
periodic orbits coincide $\gamma_{0}=\gamma_{0}^{+}=\gamma_{0}^{-}$,
then any point in $\mathcal{N}^{+}\left(\gamma_{0}\right)\cap\mathcal{N}^{-}\left(\gamma_{0}\right)$
is called a \textbf{homoclinic point} and its orbit $t\rightarrow\phi^{t}\left(m\right)$
is a \textbf{homoclinic orbit}, i.e., it tends to $\gamma_{0}$ when
$t\rightarrow\pm-\infty$.

In general, the two manifolds $\mathcal{N}^{+}\left(\gamma_{0}^{+}\right)$
and $\mathcal{N}^{-}\left(\gamma_{0}^{-}\right)$ have no reason to
coincide and the set of heteroclinic points may be very complicated. 

\begin{center}\includegraphics{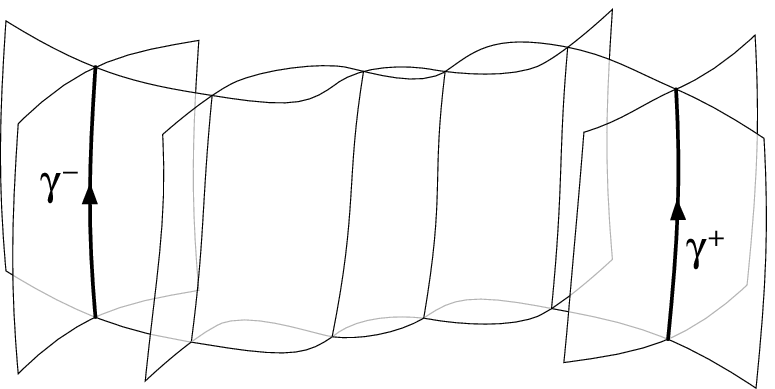} \end{center}

Nevertheless, Mel'nikov's theory deals precisely with perturbations
$H_{\varepsilon}$ of a Hamiltonian $H_{0}$ with two hyperbolic periodic
orbits $\gamma_{0}^{+}$ and $\gamma_{0}^{-}$ such that the closure
of the stable manifolds $\mathcal{N}^{+}\left(\gamma_{0}^{+}\right)$
do coincide with the closure of the unstable manifold $\mathcal{N}^{-}\left(\gamma_{0}^{-}\right)$.
We introduce the following notation.

\begin{defn}
\label{def_connected_component_stable_manifold}We define $\mathcal{N}_{0}^{\pm}:=\mathcal{N}^{\pm}\left(\gamma_{0}^{\pm}\right)$.
In the $4$-dimensional case, $\mathcal{N}^{\pm}\left(\gamma_{0}^{\pm}\right)$
denotes one connected component of the (un)stable manifold, as explained
in the previous section.
\end{defn}

\begin{defn}
\label{def_heteroclinic_homoclinic_situation}From now on, we focus
on the following two situations :
\begin{itemize}
\item \noindent \textbf{Heteroclinic situation.} We suppose that the Hamiltonian
$H_{0}$ admits two hyperbolic periodic orbits $\gamma_{0}^{+}$ and
$\gamma_{0}^{-}$. Moreover, we suppose that the closure of the stable
manifold $\mathcal{N}_{0}^{+}$ of $\gamma_{0}^{+}$ coincides with
the closure of the unstable manifold $\mathcal{N}_{0}^{-}$ of $\gamma_{0}^{-}$,
and we denote by $\mathcal{N}_{0}=\overline{\mathcal{N}_{0}^{+}}=\overline{\mathcal{N}_{0}^{-}}$
this heteroclinic manifold.
\item \noindent \textbf{Homoclinic situation.} We suppose that the Hamiltonian
$H_{0}$ admits one hyperbolic periodic orbit $\gamma_{0}$. Moreover,
we suppose that the closures of its stable and unstable manifolds
coincide, and we denote by $\mathcal{N}_{0}=\overline{\mathcal{N}_{0}^{+}}=\overline{\mathcal{N}_{0}^{-}}$
this homoclinic manifold.
\end{itemize}
\end{defn}
Despite in general the manifolds $\mathcal{N}_{0}^{+}$ and $\mathcal{N}_{0}^{-}$
are immersed in $\mathcal{M}$ in a complicated way, when $\overline{\mathcal{N}_{0}^{+}}$
and $\overline{\mathcal{N}_{0}^{-}}$ coincide, then they have the
following nice form. 

\begin{lem}
\label{lem_heteroclinic_manifold_nice}There is a time $\kappa$ such
that $\mathcal{N}_{0,\kappa}^{+}\cup\mathcal{N}_{0,\kappa}^{-}=\mathcal{N}_{0}$,
where $\mathcal{N}_{0,\kappa}^{\pm}$ denote the manifolds defined
in Definition \ref{def_stable_manifold_T}. Moreover, one has \[
\lim_{\kappa\rightarrow+\infty}\mathcal{N}_{0,\kappa}^{+}\cap\mathcal{N}_{0,\kappa}^{-}=\mathcal{N}_{0}.\]

\end{lem}
One the picture below, the thin lines represent $\mathcal{N}_{0}$
while the thick ones represent $\mathcal{N}_{0,\kappa}^{\pm}$.

\begin{center}\begin{tabular}{|c|c|c|}
\cline{2-2} \cline{3-3} 
\multicolumn{1}{c|}{}&
\multicolumn{1}{c|}{Heteroclinic}&
\multicolumn{1}{c|}{Homoclinic}\tabularnewline
\hline 
\multicolumn{1}{|c|}{small $\kappa$}&
\includegraphics[%
  scale=0.5]{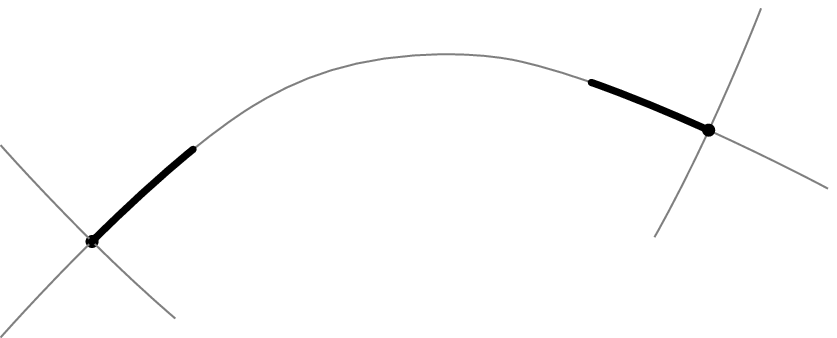}&
\includegraphics[%
  scale=0.5]{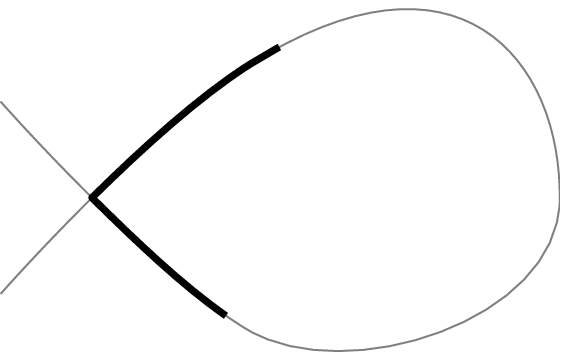}\tabularnewline
\hline 
large $\kappa$&
\includegraphics[%
  scale=0.5]{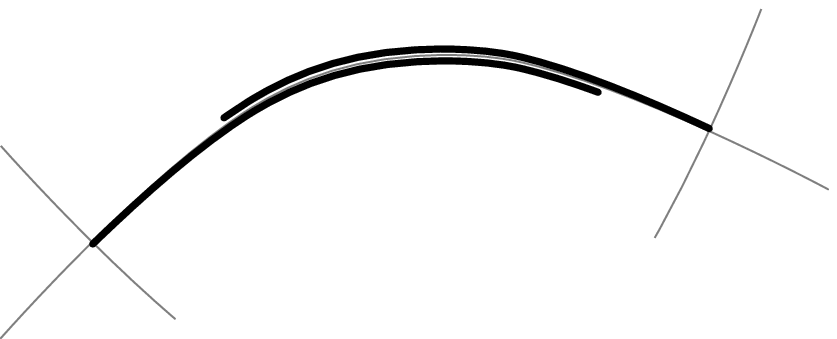}&
\includegraphics[%
  scale=0.5]{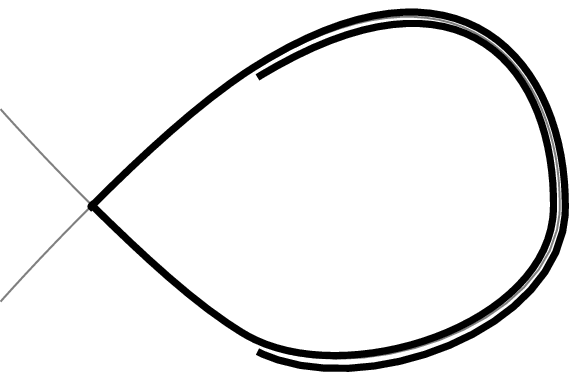}\tabularnewline
\hline
\end{tabular}\end{center}

\subsection{Heteroclinic/homoclinic orbits and the Mel'nikov $1$-form}

\subsubsection{Splitting of heteroclinic/homoclinic Lagrangian submanifolds and
the Mel'nikov $1$-form\label{sec_heteroclinic_melnikov}}

Let $H_{0}\in C^{\infty}\left(\mathcal{M}\right)$ be a Hamiltonian
either in the heteroclinic or in the homoclinic situation (see Definition
\ref{def_heteroclinic_homoclinic_situation}) and let $\mathcal{N}_{0,\kappa}^{\pm}$
be the corresponding manifolds for a chosen large $\kappa>0$. A very
important consequence of the hyperbolicity of the periodic orbits
$\gamma_{0}^{\pm}$ is that the system is structurally stable \cite{fenichel}.
This means that if $H_{\varepsilon}\in C^{\infty}\left(\mathcal{M}\right)$
is a perturbation of $H_{0}$, then in a neighbourhood of $\gamma_{0}^{\pm}$
there is an hyperbolic periodic orbit $\gamma_{\varepsilon}^{\pm}$
of $H_{\varepsilon}$ $\varepsilon$-close to $\gamma_{0}^{\pm}$.
Moreover, the stable and unstable manifolds of $H_{\varepsilon}$
are $\varepsilon$-close to those of $\gamma_{\varepsilon}^{\pm}$.
Actually, the smoothness of $H_{\varepsilon}$ with respect to $\varepsilon$
implies the smoothness of $\gamma_{\varepsilon}^{\pm}$ and $\mathcal{N}_{\varepsilon}^{\pm}$.

Restricting to a compact part as in Definition \ref{def_stable_manifold_T},
we thus have two families of periodic orbits $\gamma_{\varepsilon}^{\pm}$
together with two families of manifolds $\mathcal{N}_{\varepsilon,\kappa}^{\pm}$
and we want to detect at first order in $\varepsilon$ the intersections
$\mathcal{N}_{\varepsilon,\kappa}^{+}\cap\mathcal{N}_{\varepsilon,\kappa}^{-}$
for small $\varepsilon$ using the Mel'nikov $1$-form defined in
the first section. Unfortunately, we are not strictly speaking in
the Mel'nikov setting since $\mathcal{N}_{0,\kappa}^{+}$ and $\mathcal{N}_{0,\kappa}^{-}$
do not coincide exactly. Nevertheless, for large $\kappa$ the intersection
$\mathcal{N}_{0,\kappa}^{+}\cap\mathcal{N}_{0,\kappa}^{-}$ tends
to $\mathcal{N}_{0}$. In order to avoid an useless complexification
of the notations, we will make a slight misuse of notations by using
the Mel'nikov $1$-form $\beta\in\Omega^{1}\left(\mathcal{N}_{0}\right)$
for the families $\mathcal{N}_{\varepsilon,\kappa}^{\pm}$, being
implicitly understood that it is defined only inside the intersection
$\mathcal{N}_{0,\kappa}^{+}\cap\mathcal{N}_{0,\kappa}^{-}$, i.e.,
away from the periodic orbits $\gamma_{0}^{\pm}$.

Since the system is Hamiltonian, the families $\mathcal{N}_{\varepsilon,\kappa}^{\pm}$
are included in a level set $H_{\varepsilon}=cst\left(\varepsilon\right)$
for each $\varepsilon$ and we are thus in the constrained setting
developed in Section \ref{sec_lagrangian_with_constraints}. We know
from there that the transverse infinitesimal intersections of $\mathcal{N}_{\varepsilon,\kappa}^{+}$
and $\mathcal{N}_{\varepsilon,\kappa}^{-}$ are slightly deformed
by the perturbation, and this shows that the Mel'nikov $1$-form is
the right object for detecting the existence of some of the intersections
of $\mathcal{N}_{\varepsilon,\kappa}^{+}$ and $\mathcal{N}_{\varepsilon,\kappa}^{-}$
when $\varepsilon\neq0$, i.e., heteroclinic points between $\gamma_{\varepsilon}^{-}$
and $\gamma_{\varepsilon}^{+}$.

\subsubsection{An invariance property of the Mel'nikov $1$- form }

Let $H_{\varepsilon}\in C^{\infty}\left(\mathcal{M}\right)$ be a
perturbation of a Hamiltonian $H_{0}\in C^{\infty}\left(\mathcal{M}\right)$
either in the heteroclinic or in the homoclinic situation, and let
$\beta\in\Omega^{1}\left(\mathcal{N}_{0}\right)$ be the associated
Mel'nikov $1$-form. Lemma \ref{lem_beta_evaluee_sur_XH0} says that
$\beta\left(X_{H_{0}}\right)=0$ everywhere on $\mathcal{N}_{0}$.
This implies that the zeros of the Mel'nikov $1$-form come together
with their orbit, as explained below.

\begin{lem}
If $m\in\mathcal{N}_{0}$ is an infinitesimal intersection, $\beta_{m}=0$,
then each point of the orbit $\phi_{X_{H_{0}}}^{t}\left(m\right)$
is so. If $m$ is transversal in the constraint then each point of
the orbit $\phi_{X_{H_{0}}}^{t}\left(m\right)$ is so.
\end{lem}
\begin{proof}
The first point comes directly from the Cartan's formula $\mathcal{L}_{X_{H_{0}}}\beta=X_{H_{0}}\lrcorner d\beta+d\left(\beta\left(X_{H_{0}}\right)\right)$.
The first terms vanishes since $\beta$ is closed and the second one
vanishes thanks to Lemma \ref{lem_beta_evaluee_sur_XH0}. The Mel'nikov
$1$-form is thus invariant by the flow of $X_{H_{0}}$ and the first
point is proved. To prove the second one, let us choose an affine
connection $\nabla$ such that $\nabla X_{H_{0}}=0$ in the neighbourhood
$\mathcal{O}$ of a transversal infinitesimal intersection $m$. This
is always possible since $X_{H_{0}}$ does not vanish on $\mathcal{N}_{0}$.
We will show that $\mathcal{L}_{X_{H_{0}}}\left(\nabla\beta\right)=0$
and this will prove the second point. To evaluate $\mathcal{L}_{X_{H_{0}}}\left(\nabla\beta\right)\left(Y,Z\right)$
at a point $m$, we extend $Y$ and $Z$ to $\mathcal{O}$ in such
a way that $\nabla Y=0$ and $\nabla Z=0$. Since $X_{H_{0}}$, $Y$
and $Z$ are parallel vector fields, they commute with each other.
This implies that \[
\mathcal{L}_{X_{H_{0}}}\left(\nabla\beta\right)\left(Y,Z\right)=\mathcal{L}_{X_{H_{0}}}\left(\left(\nabla\beta\right)\left(Y,Z\right)\right)=\mathcal{L}_{X_{H_{0}}}\left(Y\left(\beta\left(Z\right)\right)\right).\]
The Leibniz rule for the Lie derivative then gives \[
\mathcal{L}_{X_{H_{0}}}\left(Y\left(\beta\left(Z\right)\right)\right)=Y\left(\left(\mathcal{L}_{X_{H_{0}}}\beta\right)\left(Z\right)\right)\]
 and this vanishes as we have shown earlier. The $\left(2,0\right)$-tensor
field $\nabla\beta$ is thus invariant by the flow of $X_{H_{0}}$.
According to Proposition \ref{prop_constrained_lagrangian_transverse_condition},
a point $m$ is a transversal infinitesimal intersection, iff $\ker\nabla\beta$
is exactly the line generated by $X_{H_{0}}$. Now, since $X_{H_{0}}$
and $\nabla\beta$ are invariant by the flow of $X_{H_{0}}$, then
we have \[
\left(\phi_{X_{H_{0}}}^{-t}\right)^{*}\left(X_{H_{0}}\lrcorner\nabla\beta\right)_{m}=\left(X_{H_{0}}\lrcorner\nabla\beta\right)_{\phi_{X_{H_{0}}}^{t}\left(m\right)}.\]
This means that $\ker\nabla\beta$ at $m$ is generated by $X_{H_{0}}$
iff it is so at each point of the orbit $\phi_{X_{H_{0}}}^{t}\left(m\right)$.
\end{proof}

\subsubsection{Mel'nikov potentials\label{sub_Melnikov_potentials}}

\begin{lem}
The Mel'ni{\-}kov $1$-form $\beta$ is exact. Any primitive, i.e.,
any function $L\in C^{\infty}\left(\mathcal{N}_{0}\right)$ with $\beta=dL$,
is called a Mel'ni{\-}kov potential.
\end{lem}
\begin{proof}
We already know from Lemma \ref{lem_melnikov_is_closed} that $\beta$
is closed. Therefore, it is exact if $\int_{\gamma}\beta=0$ for cycles
$\gamma$ generating the homology group $H_{1}\left(\mathcal{N}_{0}\right)$.
Actually, we see from Definition \ref{def_melnikov_1form} that $\beta$
is a difference $\beta=\beta^{+}-\beta^{-}$, where the $\beta^{\pm}$
are closed $1$-forms defined on $\mathcal{N}_{0}\backslash\gamma_{0}^{\mp}$.
Now, the manifolds $\mathcal{N}_{0}\backslash\gamma_{0}^{\mp}$ are
diffeomorphic to $S^{1}\times\mathbb{R}^{d-1}$ (or $S^{1}\times\mathbb{R}^{+}$
for some $4$-dimensional systems, as explained at the end of Section
\ref{sub_moebius}). Their homology is thus generated precisely by
the cycle $\gamma_{0}^{\pm}$. But these are trajectories of $H_{0}$.
Therefore, one has \[
\int_{\gamma_{0}^{\pm}}\beta^{\pm}=\frac{1}{\tau^{\pm}}\int_{0}^{\tau^{\pm}}\beta^{\pm}\left(X_{H_{0}}\right)\circ\phi_{X_{H_{0}}}^{s}\left(m_{0}\right)\, ds,\]
with $m_{0}$ any point on $\gamma_{0}^{\pm}$ and $\tau^{\pm}$ the
period of the orbit $\gamma_{0}^{\pm}$. Using Lemma \ref{lem_beta_evaluee_sur_XH0},
we conclude that $\int_{\gamma_{0}^{\pm}}\beta=0$ and therefore $\beta$
is exact.
\end{proof}
Despite this apparently pleasant property, we will not use Mel'ni{\-}kov
potentials, for several reasons. First of all, the object which parameterises
the deformations of the Lagrangian (stable and unstable) submanifolds
is really a closed $1$-form and not its primitive. Second, the heteroclinic
points are detected by the zeros of $\beta$, i.e., the critical points
of a primitive $L$. Thus, in any case, one has to compute the derivative
of $L$. Third, it might happen that $\beta$ admits a nice integral
expression, but $L$ does not, as we explain later in Section \ref{sec_remark_on_potentials}.

\subsection{Integral expression in the CI Case}

Consider a Hamiltonian $H_{0}\in C^{\infty}\left(\mathcal{M}\right)$
either in the heteroclinic or in the homoclinic situation and let
$H_{\varepsilon}\in C^{\infty}\left(\mathcal{M}\right)$ be a perturbation.
The definition of the Mel'nikov $1$-form associated with the deformed
stable and unstable manifolds does actually not take into account
the dynamical character of these manifolds. But, we will now show
that there is an integral expression for the contraction $\beta\left(X_{A}\right)$,
when $A$ is any conserved quantity, i.e., a function on $\mathcal{M}$
satisfying $\left\{ A,H_{0}\right\} =0$. 

When such a conserved quantity $A$ exists, the dynamical character
of the system allows one to give an integral expression for $\beta\left(X_{A}\right)$,
which corresponds in some special cases to the object called \emph{Mel'nikov
function} presented \emph{}in the literature. Unfortunately, in the
general case this integral does not converge and one has to give a
prescription to make it converge. We explain this issue in Section
\ref{sec_integral_prescription}. Nevertheless, there are two cases
where the integral converges. They are discussed in Sections \ref{sec_perturbation_critical_on_orbits}
and \ref{sec_good_A}. First, we describe the situation where the
perturbation is critical on the orbits $\gamma_{0}^{+}$ and $\gamma_{0}^{-}$.
Finally, we consider the case when the conserved quantity $A$ is
critical on $\gamma_{0}^{+}$ and $\gamma_{0}^{-}$. We notice that
in order to describe completely $\beta$, one needs to have $d$ Hamiltonian
vector fields $X_{A_{1}}$,..., $X_{A_{d}}$ tangent to $\mathcal{N}_{0}$
and linearly independent. This arises precisely when $H_{0}$ is completely
integrable and the $A_{j}$'s are the components of a momentum map.
In Section \ref{sec_good_A}, we explain how many linearly independent
$A_{j}$'s critical on the orbits $\gamma_{0}^{+}$ and $\gamma_{0}^{-}$
one can have.

\subsubsection{Momentum maps in presence of transversally hyperbolic periodic orbits}

The presence of a hyperbolic periodic orbit for a Hamiltonian $H$
implies certain properties for its conserved quantities $A$, $\left\{ A,H\right\} =0$,
as follows. 

\begin{prop}
\label{prop_commutant_avec_orbit_hyperbolic}Suppose $H$ has a transversally
hyperbolic periodic orbit $\gamma$. Then, each conserved quantity
$A$ is constant on the stable and unstable manifolds $\mathcal{N}^{\pm}\left(\gamma\right)$,
i.e., $A\left(\mathcal{N}^{+}\left(\gamma\right)\right)=A\left(\mathcal{N}^{-}\left(\gamma\right)\right)=A\left(\gamma\right)$,
and its vector field $X_{A}$ is tangent to $\mathcal{N}^{+}\left(\gamma\right)$
and $\mathcal{N}^{-}\left(\gamma\right)$. Moreover, there is a constant
$c\left(A\right)$ such that $X_{A}=c\left(A\right)X_{H}$ at each
point of $\gamma$.
\end{prop}
\begin{proof}
The commutation relation $\left\{ A,H\right\} =0$ implies that the
orbits of $X_{H}$ are included in the level sets $\left\{ m,A\left(m\right)=a\right\} $,
$a\in\mathbb{R}$. In particular one has $\gamma\subset\left\{ m,A\left(m\right)=a\right\} $
for $a$. Moreover, by definition, for each point $m$ on the stable
manifold $\mathcal{N}^{+}\left(\gamma\right)$, one has $\phi_{X_{H}}^{t}\left(m\right)\rightarrow\gamma$,
when $t\rightarrow+\infty$. Since the function $A$ is constant on
the trajectories of $X_{H}$, we must have $A\left(m\right)=A\circ\phi_{X_{H}}^{t}\left(m\right)$
and the limit $t\rightarrow+\infty$ yields $A\left(m\right)=A\left(\gamma\right)$
for each $m\in\mathcal{N}^{+}\left(\gamma\right)$, i.e., $A$ is
constant on $\mathcal{N}^{+}\left(\gamma\right)$. A similar argument
shows that $A$ is also constant on $\mathcal{N}^{-}\left(\gamma\right)$.
Since $\mathcal{N}^{\pm}\left(\gamma\right)$ is Lagrangian, the inclusion
$T_{m}\mathcal{N}^{\pm}\left(\gamma\right)\subset\ker dA_{m}$ at
the point $m\in\mathcal{N}^{\pm}\left(\gamma\right)$ is equivalent
by duality to $X_{A}\in T_{m}\mathcal{N}^{\pm}\left(\gamma\right)$.
Now, at each point $m\in\gamma$, the intersection $T_{m}\mathcal{N}^{-}\left(\gamma\right)\cap T_{m}\mathcal{N}^{+}\left(\gamma\right)$
is exactly $T_{m}\gamma$ and therefore $X_{A}\in T_{m}\gamma$. Consequently,
there is a function $c:\gamma\rightarrow\mathbb{R}$ such that one
has the relation $X_{A}=c\left(m\right)X_{H}$, at $m\in\gamma$ .
Moreover, $X_{A}$ is invariant under the flow of $X_{H}$, since
$\left\{ H,A\right\} =0$ and thus $\mathcal{L}_{X_{H}}X_{A}=0$.
This implies that $c\left(m\right)$ is independent of $m$.
\end{proof}
According to Proposition \ref{prop_commutant_avec_orbit_hyperbolic},
if $H_{0}$ admits a momentum map $\mathbf{A}=\left(A_{1},...,A_{d}\right)$
which is regular on $\mathcal{N}_{0}\setminus\left(\gamma_{0}^{+}\cup\gamma_{0}^{-}\right)$,
then the Hamiltonian vector fields $X_{A_{1}},...,X_{A_{d}}$ form
a basis of $T_{m}\mathcal{N}_{0}$ at each point $m\in\mathcal{N}_{0}\setminus\left(\gamma_{0}^{+}\cup\gamma_{0}^{-}\right)$.
Therefore, the Mel'nikov $1$-form $\beta\in\Omega^{1}\left(\mathcal{N}_{0}\right)$
associated to any perturbation $H_{\varepsilon}$ is fully understood
whenever one is able to compute the evaluations $\beta\left(X_{A_{1}}\right),...,\beta\left(X_{A_{d}}\right)$.

\subsubsection{Integral expression with prescription\label{sec_integral_prescription}}

Thanks to Proposition \ref{prop_commutant_avec_orbit_hyperbolic},
the vector field $X_{A}$ associated to any conserved quantity $A$
is tangent to the heteroclinic/homoclinic manifold $\mathcal{N}_{0}$,
and one can thus evaluate the Mel'nikov $1$-form $\beta$ on it.
This evaluation can express in terms of an integral involving the
first order perturbation $H_{1}=\left.\frac{dH_{\varepsilon}}{d\varepsilon}\right|_{\varepsilon=0}$,
as the next theorem shows. 

\begin{thm}
\label{theo_integral_prescription}Let $H_{0}\in C^{\infty}\left(\mathcal{M}\right)$
be a Hamiltonian either in the heteroclinic or in the homoclinic situation.
Let $H_{\varepsilon}\in C^{\infty}\left(\mathcal{M}\right)$ be a
perturbed Hamiltonian and $\beta\in\Omega^{1}\left(\mathcal{N}_{0}\right)$
the associated Mel'nikov $1$-form. Then, for any conserved quantity
$A\in C^{\infty}\left(\mathcal{M}\right)$ and any point $m\in\mathcal{N}_{0}\setminus\left(\gamma_{0}^{+}\cup\gamma_{0}^{-}\right)$,
one has the following formula \[
\beta\left(X_{A}\right)_{m}=\frac{d}{d\varepsilon}\left.A\left(m_{\varepsilon}^{+}\right)\right|_{\varepsilon=0}-\frac{d}{d\varepsilon}\left.A\left(m_{\varepsilon}^{-}\right)\right|_{\varepsilon=0}+\lim_{n\rightarrow\infty}\int_{-n\tau_{0}^{-}}^{n\tau_{0}^{+}}\left\{ H_{1},A\right\} \circ\phi_{X_{H_{0}}}^{t}\left(m\right)dt,\]
where $\tau_{\varepsilon}^{\pm}$ is the period of $\gamma_{\varepsilon}^{\pm}$
and the point $m_{\varepsilon}^{\pm}\in\gamma_{\varepsilon}^{\pm}$
is given by \[
m_{\varepsilon}^{\pm}=\lim_{n\rightarrow\infty}\phi_{X_{H_{\varepsilon}}}^{\mp n\tau_{\varepsilon}^{\pm}}\circ\phi_{X_{\varepsilon}^{\pm}}^{\varepsilon}\left(m\right).\]

\end{thm}
\begin{center}\includegraphics{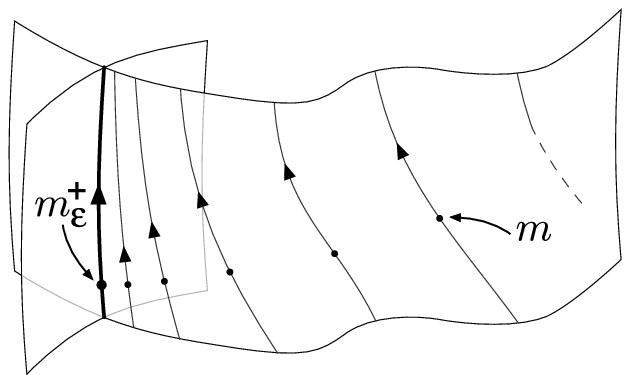}\end{center}

\begin{proof}
Fix a constant $\kappa$ large enough for $m$ to belong to $\mathcal{N}_{0,\kappa}^{+}\cap\mathcal{N}_{0,\kappa}^{-}$.
Remember that the Mel'nikov $1$-form is given in Definition \ref{def_melnikov_1form}
by $\beta=\iota^{*}\left(\left(X_{0}^{+}-X_{0}^{-}\right)\lrcorner\omega\right)$,
where $X_{\varepsilon}^{+}$ (resp. $X_{\varepsilon}^{-}$) generates
the stable (resp. unstable) manifold $\mathcal{N}_{\varepsilon,\kappa}^{+}$
(resp. $\mathcal{N}_{\varepsilon,\kappa}^{-}$) of the orbit $\gamma_{\varepsilon}^{+}$
(resp. $\gamma_{\varepsilon}^{-}$). For the evaluation on $X_{A}$,
we have to compute both terms $\omega\left(X_{0}^{\pm},X_{A}\right)$,
which are nothing but $X_{0}^{\pm}\left(A\right)$, i.e., $\left.\frac{d}{d\varepsilon}A\circ\phi_{X_{\varepsilon}^{\pm}}^{\varepsilon}\left(m\right)\right|_{\varepsilon=0}$.
Now, by definition of $X_{\varepsilon}^{\pm}$, the point $\phi_{X_{\varepsilon}^{\pm}}^{\varepsilon}\left(m\right)$
is on $\mathcal{N}_{\varepsilon,\kappa}^{\pm}$. We will compute an
expression for $A\circ\phi_{X_{\varepsilon}^{+}}^{\varepsilon}\left(m\right)$
(and later similarly for $A\circ\phi_{X_{\varepsilon}^{-}}^{\varepsilon}\left(m\right)$)
in terms of an integral which converges uniformly with respect to
$\varepsilon$ and then take the derivative. For any time $T$, one
has the relation \[
A\circ\phi_{X_{\varepsilon}^{+}}^{\varepsilon}\left(m\right)=A\circ\phi_{X_{H_{\varepsilon}}}^{T}\circ\phi_{X_{\varepsilon}^{+}}^{\varepsilon}\left(m\right)-\int_{0}^{T}X_{H_{\varepsilon}}\left(A\right)\circ\phi_{X_{H_{\varepsilon}}}^{t}\circ\phi_{X_{\varepsilon}^{+}}^{\varepsilon}\left(m\right)dt,\]
where we have used $\frac{d}{dt}A\circ\phi_{X_{H_{\varepsilon}}}^{t}=X_{H_{\varepsilon}}\left(A\right)\circ\phi_{X_{H_{\varepsilon}}}^{t}$.
Now, $\phi_{X_{\varepsilon}^{+}}^{\varepsilon}\left(m\right)$ is
on the stable manifold $\mathcal{N}_{\varepsilon}^{+}$ and $\phi_{X_{H_{\varepsilon}}}^{T}\circ\phi_{X_{\varepsilon}^{+}}^{\varepsilon}\left(m\right)$
tends to the cycle $\gamma_{\varepsilon}^{+}$ when $T\rightarrow\infty$.
If one considers the discrete times $T=n\tau_{\varepsilon}^{+}$,
with $\tau_{\varepsilon}^{+}$ the period of $\gamma_{\varepsilon}^{+}$,
then $\phi_{X_{H_{\varepsilon}}}^{n\tau_{\varepsilon}^{+}}\circ\phi_{X_{\varepsilon}^{+}}^{\varepsilon}\left(m\right)$
has a limit on $\gamma_{\varepsilon}^{+}$ when $n\rightarrow\infty$
and this limit is uniform in $\varepsilon$. Indeed, $\phi_{X_{\varepsilon}^{+}}^{\varepsilon}\left(m\right)$
is on the stable manifold of some point $m_{\varepsilon}^{+}\in\gamma_{\varepsilon}^{+}$,
i.e., $dist\left(\phi_{X_{H_{\varepsilon}}}^{T}\left(m_{\varepsilon}^{+}\right),\phi_{X_{H_{\varepsilon}}}^{T}\circ\phi_{X_{\varepsilon}^{+}}^{\varepsilon}\left(m\right)\right)\leq C_{\varepsilon}e^{-T\lambda_{\varepsilon}}$
for large $T$. Taking the maximum $C$ over $\varepsilon$ of the
constant $C_{\varepsilon}$ and the minimum $\lambda$ of the Liapounov
exponent $\lambda_{\varepsilon}$, one obtains\[
dist\left(\phi_{X_{H_{\varepsilon}}}^{T}\left(m_{\varepsilon}^{+}\right),\phi_{X_{H_{\varepsilon}}}^{T}\circ\phi_{X_{\varepsilon}^{+}}^{\varepsilon}\left(m\right)\right)\leq Ce^{-T\lambda}\]
 for all $\varepsilon$ and all $T$. Now, for $T=n\tau_{\varepsilon}^{+}$
one has $\phi_{X_{H_{\varepsilon}}}^{n\tau_{\varepsilon}^{+}}\left(m_{\varepsilon}^{+}\right)=m_{\varepsilon}^{+}$
and therefore for all positive integers $n$ and all $\varepsilon$,
one has \[
dist\left(m_{\varepsilon}^{+},\phi_{X_{H_{\varepsilon}}}^{n\tau_{\varepsilon}^{+}}\circ\phi_{X_{\varepsilon}^{+}}^{\varepsilon}\left(m\right)\right)\leq2Ce^{-n\tau\lambda},\]
where $\tau=\min_{\varepsilon}\tau_{\varepsilon}^{+}$. This shows
the uniformity with respect to $\varepsilon$ of the limit point $m_{\varepsilon}^{+}=\lim_{n\rightarrow\infty}\phi_{X_{H_{\varepsilon}}}^{n\tau_{\varepsilon}^{+}}\circ\phi_{X_{\varepsilon}^{+}}^{\varepsilon}\left(m\right)$,
which implies in return the uniformity of the limit of $\int_{0}^{n\tau_{\varepsilon}^{+}}X_{H_{\varepsilon}}\left(A\right)\circ\phi_{X_{H_{\varepsilon}}}^{t}\circ\phi_{X_{\varepsilon}^{+}}^{\varepsilon}\left(m\right)dt$.
The term $X_{0}^{+}\left(A\right)$ is thus given by the expression
\[
X_{0}^{+}\left(A\right)=\frac{d}{d\varepsilon}\left.A\left(m_{\varepsilon}^{+}\right)\right|_{\varepsilon=0}-\frac{d}{d\varepsilon}\left.\lim_{n\rightarrow\infty}\int_{0}^{n\tau_{\varepsilon}^{+}}X_{H_{\varepsilon}}\left(A\right)\circ\phi_{X_{H_{\varepsilon}}}^{t}\circ\phi_{X_{\varepsilon}^{+}}^{\varepsilon}\left(m\right)dt\right|_{\varepsilon=0}.\]
If we perform a second order Taylor expansion on $H_{\varepsilon}$
with respect to $\varepsilon$, i.e., $H_{\varepsilon}=H_{0}+\varepsilon H_{1}+\varepsilon^{2}K_{\varepsilon}$
with $K_{\varepsilon}$ depending smoothly on $\varepsilon$, then
one has \[
X_{H_{\varepsilon}}\left(A\right)=\varepsilon\left\{ A,H_{1}+\varepsilon K_{\varepsilon}\right\} \]
 since $\left\{ H_{0},A\right\} =0$. This gives \[
X_{0}^{+}\left(A\right)=\frac{d}{d\varepsilon}\left.A\left(m_{\varepsilon}^{+}\right)\right|_{\varepsilon=0}+\lim_{n\rightarrow\infty}\int_{0}^{n\tau_{0}^{+}}\left\{ H_{1},A\right\} \circ\phi_{X_{H_{0}}}^{t}\left(m\right)dt,\]
for all $m$ on $\mathcal{N}_{\varepsilon,\kappa}^{+}$. A completely
similar procedure yields the corresponding expression for the term
$X_{0}^{-}\left(A\right)$, for all $m$ on $\mathcal{N}_{\varepsilon,\kappa}^{-}$,
and we obtain the claimed expression for $\beta\left(X_{A}\right)=X_{0}^{+}\left(A\right)-X_{0}^{-}\left(A\right)$.
\end{proof}
As mentioned in Section \ref{sec_heteroclinic_melnikov}, this integral
expression is not valid in the whole $\mathcal{N}_{0}$, but rather
in $\mathcal{N}_{0,\kappa}^{+}\cap\mathcal{N}_{0,\kappa}^{y-}$ for
arbitrarily large $\kappa$. Indeed, one should keep in mind that
the convergence of the different limits in this expression becomes
worse and worse when one let $m$ get closer to $\gamma_{0}^{+}$
or $\gamma_{0}^{-}$. This reflects the so-called {}``heteroclinic
entanglement'' phenomenon that occurs near $\gamma_{0}^{+}$ and
$\gamma_{0}^{-}$.

It might seem to the reader that this integral expression is not very
easy to handle, but this is unfortunately the only one available without
any further assumptions. In the next two subsections though, we consider
special cases for which this expression takes a simpler form.

\subsubsection{Homoclinic case with a perturbation critical on the orbits\label{sec_perturbation_critical_on_orbits}}

When the perturbation is critical on the orbits $\gamma_{0}^{+}$
and $\gamma_{0}^{-}$, i.e., $d\left(H_{\varepsilon}-H_{0}\right)=0$,
then both cycles remain periodic orbits of the perturbed dynamics
$H_{\varepsilon}$ for all $\varepsilon$. This means that there is
a family of energies $E_{\varepsilon}$ such that the family of orbits
$\gamma_{\varepsilon}^{+}$, included in the energy levels $\left\{ H_{\varepsilon}=E_{\varepsilon}\right\} $,
is actually constant, i.e., $\gamma_{\varepsilon}^{+}=\gamma_{0}$.
This would yield a simplification in the formula of Theorem \ref{theo_integral_prescription}
since the term $\frac{d}{d\varepsilon}\left.A\left(m_{\varepsilon}^{+}\right)\right|_{\varepsilon=0}$
would vanish and the integral would converge without prescription
on the way to take the limit. Of course, one could do this rather
for the unstable orbit $\gamma_{\varepsilon}^{-}$, but unfortunately
it is impossible to do this simultaneously for both $\gamma_{\varepsilon}^{+}$
and $\gamma_{\varepsilon}^{-}$, except when $H_{\varepsilon}$ takes
the same value on $\gamma_{\varepsilon}^{+}$ and $\gamma_{\varepsilon}^{-}$.
In the heteroclinic situation, this must be an assumption whereas
in the homoclinic one this is automatic. Actually, one can obtain
this result assuming only that the first order perturbation $H_{1}=\left.\frac{dH_{\varepsilon}}{d\varepsilon}\right|_{\varepsilon=0}$
is critical on $\gamma_{0}^{+}$ and $\gamma_{0}^{-}$, as the next
theorem shows.

\begin{thm}
\label{theo_good_perturbation}Let $H_{0}\in C^{\infty}\left(\mathcal{M}\right)$
be a Hamiltonian either in the heteroclinic or in the homoclinic situation.
Let $H_{\varepsilon}\in C^{\infty}\left(\mathcal{M}\right)$ be a
perturbed Hamiltonian such that the first order perturbation $H_{1}$
is critical on both orbits $\gamma_{0}^{+}$ and $\gamma_{0}^{-}$.
Moreover, in the heteroclinic situation, assume that $H_{1}\left(\gamma_{0}^{+}\right)=H_{1}\left(\gamma_{0}^{-}\right)$.
Let $\gamma_{\varepsilon}^{\pm}$ be families of periodic orbits of
$H_{\varepsilon}$ included in the energy levels $\left\{ H_{\varepsilon}=E_{\varepsilon}\right\} $,
with $E_{1}=H_{1}\left(\gamma_{0}^{\pm}\right)$ and let $\beta\in\Omega^{1}\left(\mathcal{N}_{0}\right)$
be the associated Mel'nikov $1$-form. Then, for any conserved quantity
$A\in C^{\infty}\left(\mathcal{M}\right)$ and any point $m\in\mathcal{N}_{0}\setminus\left(\gamma_{0}^{+}\cup\gamma_{0}^{-}\right)$,
the following formula holds : \[
\beta\left(X_{A}\right)_{m}=\int_{-\infty}^{+\infty}\left\{ H_{1},A\right\} \circ\phi_{X_{H_{0}}}^{t}\left(m\right)dt.\]

\end{thm}
\begin{proof}
First of all, $\beta\left(X_{A}\right)_{m}$ is given by the formula
of Theorem \ref{theo_integral_prescription}. Since $m_{\varepsilon}^{+}\in\gamma_{\varepsilon}^{+}$
one must have $\phi_{X_{H_{\varepsilon}}}^{\tau_{\varepsilon}^{+}}\left(m_{\varepsilon}^{+}\right)=m_{\varepsilon}^{+}$
for all $\varepsilon$, where $\tau_{\varepsilon}^{+}$ is the period
of $\gamma_{\varepsilon}^{+}$. Let us denote by $Y\in T_{m_{0}^{+}}\mathcal{M}$
the vector tangent to the curve $m_{\varepsilon}^{+}$ at $\varepsilon=0$,
and let us prove that $Y$ is tangent to $\gamma_{0}^{+}$. For any
function $f\in C^{\infty}\left(\mathcal{M}\right)$ one has $\frac{d}{d\varepsilon}\left.f\left(m_{\varepsilon}^{+}\right)\right|_{\varepsilon=0}=Y\left(f\right)$
and the previous equality of curves provides\[
Y\left(f\right)=\frac{d}{d\varepsilon}\left.f\circ\phi_{X_{H_{0}}}^{\tau_{0}^{+}}\left(m_{\varepsilon}^{+}\right)\right|_{\varepsilon=0}+\frac{d}{d\varepsilon}\left.f\circ\phi_{X_{H_{0}}}^{\tau_{\varepsilon}^{+}}\left(m_{0}^{+}\right)\right|_{\varepsilon=0}+\frac{d}{d\varepsilon}\left.f\circ\phi_{X_{H_{\varepsilon}}}^{\tau_{0}^{+}}\left(m_{0}^{+}\right)\right|_{\varepsilon=0}.\]
The first term is simply $Y\left(f\circ\phi_{X_{H_{0}}}^{\tau_{0}^{+}}\right)$,
i.e., $\left(\left(\phi_{X_{H_{0}}}^{\tau_{0}^{+}}\right)_{*}Y\right)f$.
The second one is $\tau_{1}X_{H_{0}}\left(f\right)\circ\phi_{X_{H_{0}}}^{\tau_{0}^{+}}\left(m_{0}^{+}\right)$
which is equal to $\tau_{1}X_{H_{0}}\left(f\right)_{m_{0}}$. And
the third one is a variation of a flow whose expression is \[
\frac{d}{d\varepsilon}\left.f\circ\phi_{X_{H_{\varepsilon}}}^{\tau_{0}^{+}}\left(m_{0}^{+}\right)\right|_{\varepsilon=0}=\int_{0}^{\tau_{0}^{+}}\left(\left(\phi_{X_{H_{0}}}^{\tau_{0}^{+}}\right)_{*}X_{H_{1}}\right)_{\phi_{X_{H_{0}}}^{\tau_{0}^{+}}\left(m_{0}^{+}\right)}f\, dt.\]
But this vanishes since by hypothesis $dH_{1}=0$ at each point of
$\gamma_{0}^{+}$ and thus $X_{H_{1}}=0$ on $\gamma_{0}^{+}$. All
together, these terms give the following equation at the point $m_{0}^{+}$\[
Y=\left(\phi_{X_{H_{0}}}^{\tau_{0}^{+}}\right)_{*}Y+\tau_{1}X_{H_{0}}.\]
Now, remember that $\left(\phi_{X_{H_{0}}}^{\tau_{0}^{+}}\right)_{*}$
at the point $m_{0}^{+}$ has the eigenvalue $1$ with multiplicity
$2$, whose eigenspace contains $X_{H_{0}}$. Decompose $Y$ accordingly,
i.e., as $Y=Y_{1}+Y_{2}$, with $\left(\phi_{X_{H_{0}}}^{\tau_{0}^{+}}\right)_{*}Y_{1}=Y_{1}$
and $Y_{2}$ in the sum of the other eigenspaces. Therefore, the component
$Y_{1}$ satisfies $Y_{1}=Y_{1}+\tau_{1}X_{H_{0}}$ which proves that
$\tau_{1}=0$ and that $\left(\phi_{X_{H_{0}}}^{\tau_{0}^{+}}\right)_{*}Y=Y$.
On the other hand, differentiating the relation $H_{\varepsilon}\left(m_{\varepsilon}^{+}\right)=E_{\varepsilon}$
with respect to $\varepsilon$ yields $Y\left(H_{0}\right)+H_{1}\left(m_{0}^{+}\right)=E_{1}$
and the hypothesis $E_{1}=H_{1}\left(\gamma_{0}^{\pm}\right)$ implies
that $Y\left(H_{0}\right)=0$. Together with the fact that $Y$ is
an eigenvector of $\left(\phi_{X_{H_{0}}}^{\tau_{0}^{+}}\right)_{*}$
with eigenvalue $1$, this shows that $Y$ is collinear to $X_{H_{0}}$,
i.e., tangent to $\gamma_{0}^{+}$. Arguing exactly in the same way,
we show that $Y$ is also tangent to the second orbit $\gamma_{0}^{-}$.
This shows that, in the formula of Theorem \ref{theo_integral_prescription},
both terms $\frac{d}{d\varepsilon}\left.A\left(m_{\varepsilon}^{+}\right)\right|_{\varepsilon=0}$
and $\frac{d}{d\varepsilon}\left.A\left(m_{\varepsilon}^{-}\right)\right|_{\varepsilon=0}$
vanish.

On the other hand, the Poisson bracket $\left\{ H_{1},A\right\} $
vanishes on the orbits $\gamma_{0}^{+}$ and $\gamma_{0}^{-}$ since
$dH_{1}$ does. This implies that the integral $\int_{-T}^{T}\left\{ H_{1},A\right\} \circ\phi_{X_{H_{0}}}^{t}\left(m\right)\, dt$
converges when $T\rightarrow\infty$ and one can replace $\lim_{n\rightarrow\infty}\int_{-n\tau_{0}^{-}}^{n\tau_{0}^{+}}$
by $\int_{-\infty}^{+\infty}$. 
\end{proof}

\subsubsection{The shrewd choice of the conserved quantity\label{sec_good_A}}

Suppose now that the perturbation does not have any special properties.
We first show that when the conserved quantity $A$ is critical on
the periodic orbits, then the expression of Theorem \ref{theo_integral_prescription}
simplifies as in Theorem \ref{theo_good_perturbation}. This result
is proved in Theorem \ref{theo_good_A} and then, we explain how many
$A$'s with this property one can have.

\begin{thm}
\label{theo_good_A}Let $H_{0}\in C^{\infty}\left(\mathcal{M}\right)$
be a Hamiltonian either in the heteroclinic or in the homoclinic situation.
Let $H_{\varepsilon}\in C^{\infty}\left(\mathcal{M}\right)$ be a
perturbed Hamiltonian. Let $\gamma_{\varepsilon}^{\pm}$ be families
of periodic orbits of $H_{\varepsilon}$ and let $\beta\in\Omega^{1}\left(\mathcal{N}_{0}\right)$
be the associated Mel'nikov $1$-form. Then, for any point $m\in\mathcal{N}_{0}\setminus\left(\gamma_{0}^{+}\cup\gamma_{0}^{-}\right)$
and any conserved quantity $A\in C^{\infty}\left(\mathcal{M}\right)$
which is critical on both orbits $\gamma_{0}^{+}$ and $\gamma_{0}^{-}$
, one has the following formula \[
\beta\left(X_{A}\right)_{m}=\int_{-\infty}^{+\infty}\left\{ H_{1},A\right\} \circ\phi_{X_{H_{0}}}^{t}\left(m\right)dt.\]

\end{thm}
\begin{proof}
We start from the expression given in Theorem \ref{theo_integral_prescription}.
The vanishing of $dA$ on $\gamma_{0}^{+}$ and $\gamma_{0}^{-}$
implies that $\left\{ H_{1},A\right\} $ vanishes on $\gamma_{0}^{+}$
and $\gamma_{0}^{-}$ too. Therefore, the integral $\int_{-T}^{T}\left\{ H_{1},A\right\} \circ\phi_{X_{H_{0}}}^{t}\left(m\right)\, dt$
converges when $T\rightarrow\infty$ and we have $\lim_{n\rightarrow\infty}\int_{-n\tau_{0}^{-}}^{n\tau_{0}^{+}}=\int_{-\infty}^{+\infty}$.
Moreover we have obviously $\frac{d}{d\varepsilon}\left.A\left(m_{\varepsilon}^{+}\right)\right|_{\varepsilon=0}=0$
and $\frac{d}{d\varepsilon}\left.A\left(m_{\varepsilon}^{-}\right)\right|_{\varepsilon=0}=0$,
and this provides the claimed expression. 
\end{proof}
Let us now address the issue of counting how many such conserved quantities
with this property one can have. Remark, that we need only $d-1$
independent $A_{j}$'s in order to describe completely $\beta$, thru
the evaluations $\beta\left(X_{A_{j}}\right)$, since $H_{0}$ itself
is a conserved quantity and we know already from Lemma \ref{lem_beta_evaluee_sur_XH0}
that $\beta\left(X_{H_{0}}\right)=0$. 

\begin{prop}
\label{prop_nombre_A_critic_sur_gamma}Suppose $H_{0}$ admits a momentum
map regular on $\mathcal{N}_{0}\setminus\left(\gamma_{0}^{+}\cup\gamma_{0}^{-}\right)$
and define $p$ by \[
p=\left\{ \begin{array}{c}
d-1\textrm{ in the homoclinic situation}\\
d-2\textrm{ in the heteroclinic situation}.\end{array}\right.\]
Then, there exist $p$ commuting constants of the motion $B_{1},...,B_{p}$
which are critical on both $\gamma_{0}^{+}$ and $\gamma_{0}^{-}$,
and satisfy\[
dH_{0}\wedge dB_{1}\wedge...\wedge dB_{p}\neq0\textrm{ on }\mathcal{N}_{0}\setminus\left(\gamma_{0}^{+}\cup\gamma_{0}^{-}\right).\]

\end{prop}
\begin{proof}
First, let us define the $d$-dimensional vector space $E\subset C^{\infty}\left(\mathcal{M}\right)$
generated by the components $\left(A_{1},...,A_{d}\right)$ of the
momentum map. Proposition \ref{prop_commutant_avec_orbit_hyperbolic}
implies that for each $A\in E$ there is a real number $c^{\pm}\left(A\right)$
such that $X_{A}-c^{\pm}\left(A\right)X_{H}$ vanishes on the orbit
$\gamma_{0}^{\pm}$. Any function $A\in E$ is critical on $\gamma_{0}^{\pm}$
precisely when $c^{\pm}\left(A\right)=0$. Actually, the map $A\rightarrow c^{\pm}\left(A\right)$
is linear with respect to $A\in E$. Indeed, one has $X_{A+A^{'}}=X_{A}+X_{A^{'}}$
which equals to $\left(c^{\pm}\left(A\right)+c^{\pm}\left(A^{'}\right)\right)X_{H}$
on the orbit $\gamma_{0}^{\pm}$, and similarly, for any constant
$\lambda$ one has $X_{\lambda A}=\lambda X_{A}$ which equals to
$\lambda c^{\pm}\left(A\right)X_{H}$ on $\gamma_{0}^{\pm}$. Moreover,
this map is non trivial. Indeed, since the $X_{A_{1}},...,X_{A_{d}}$
form a basis of the tangent $T_{m}\mathcal{N}_{0}$ at each point
$m\in\mathcal{N}_{0}\setminus\left(\gamma_{0}^{+}\cup\gamma_{0}^{-}\right)$,
the vector field $X_{H}$ restricted to $\mathcal{N}_{0}\setminus\left(\gamma_{0}^{+}\cup\gamma_{0}^{-}\right)$
is of the form $X_{H}=\Sigma_{j}a_{j}X_{A_{j}}$, with $a_{j}\in C^{^{\infty}}\left(\mathcal{N}_{0}\setminus\left(\gamma_{0}^{+}\cup\gamma_{0}^{-}\right)\right)$.
Now, since the $X_{A_{j}}$'s commute with each other and with $X_{H}$,
this implies that the functions $a_{j}$ are constant on $\mathcal{N}_{0}\setminus\left(\gamma_{0}^{+}\cup\gamma_{0}^{-}\right)$.
Therefore, $X_{H}$ coincides with $X_{A}$ on $\mathcal{N}_{0}\setminus\left(\gamma_{0}^{+}\cup\gamma_{0}^{-}\right)$,
where $A\in E$ is given by $A=\Sigma_{j}a_{j}A_{j}$. By continuity,
they coincide on the whole $\mathcal{N}_{0}$. For this $A$, the
map $c^{\pm}$ thus gives $1$. Therefore, the set $\left(c^{\pm}\right)^{-1}\left(0\right)$
is a $d-1$ dimensional hyperplane in $E$ composed of first integrals
$A$ which are critical on $\gamma_{0}^{\pm}$. In the heteroclinic
case, the hyperplanes $\left(c^{+}\right)^{-1}\left(0\right)$ and
$\left(c^{-}\right)^{-1}\left(0\right)$ do generically not coincide
and therefore intersect along a $d-2$ dimensional plane in $E$.
\end{proof}
Unfortunately, in the heteroclinic case, one can not avoid that only
$d-2$ functions $B_{j}$ are provided by Proposition \ref{prop_nombre_A_critic_sur_gamma}.
The systems usually presented in the literature (2-dimensional time-periodic)
are particular in this regard, because the flow of $H_{0}$ has the
same period on $\gamma_{0}^{+}$ and $\gamma_{0}^{-}$. In that case,
there is indeed $d-1$ functions $B_{j}$, as the next proposition
shows.

\begin{prop}
\label{prop_nombre_A_critic_sur_gamma_period_egale}Suppose $H_{0}$
admits a momentum map regular on $\mathcal{N}_{0}\setminus\left(\gamma_{0}^{+}\cup\gamma_{0}^{-}\right)$.
Assume that the periods of $\gamma_{0}^{+}$ and $\gamma_{0}^{-}$
are equal, then there exist $d-1$ commuting constants of the motion
$B_{1},...,B_{p}$ which are critical on both $\gamma_{0}^{+}$ and
$\gamma_{0}^{-}$, and satisfy\[
dH_{0}\wedge dB_{1}\wedge...\wedge dB_{p}\neq0\textrm{ on }\mathcal{N}_{0}\setminus\left(\gamma_{0}^{+}\cup\gamma_{0}^{-}\right).\]

\end{prop}
\begin{proof}
First, let $\tau_{0}^{\pm}$ be the period of $\gamma_{0}^{\pm}$.
We recall that the stable manifold $\mathcal{N}^{+}\left(\gamma_{0}^{+}\right)$
of the orbit $\gamma_{0}^{+}$ is fibred by the stable manifolds $\mathcal{N}^{+}\left(m^{+}\right)$
of all the points $m^{+}\in\gamma_{0}^{+}$, i.e., for each $m\in\mathcal{N}^{+}\left(\gamma_{0}^{+}\right)$,
the sequence $\phi_{X_{H_{0}}}^{n\tau_{0}^{+}}\left(m\right)$ tends
to a point $m^{+}\in\gamma_{0}^{+}$ when $n\rightarrow+\infty$.
Moreover, the map $m\rightarrow m^{+}$ is smooth, i.e., the limit
$\pi=\lim_{n}\phi_{X_{H_{0}}}^{n\tau_{0}^{+}}$ acts as a projection.
Of course, the same holds for the unstable manifold $\mathcal{N}^{-}\left(\gamma_{0}^{-}\right)$
of the orbit $\gamma_{0}^{-}$ .

On the other hand, if $A\in C^{\infty}\left(\mathcal{M}\right)$ is
any conserved quantity, then $\left(\phi_{X_{H_{0}}}^{n\tau_{0}^{+}}\right)_{*}X_{A}=X_{A}$
for all $n$. Therefore, $X_{A}$ is tangent to $\mathcal{N}^{+}\left(m^{+}\right)$
at a point $m$ iff $X_{A}$ vanishes at $m^{+}$ and thus on the
whole $\gamma_{0}^{+}$. Now, if the periods $\tau_{0}^{+}$ and $\tau_{0}^{-}$
are equal, then for each $m^{+}\in\gamma_{0}^{+}$ there is a point
$m^{-}\in\gamma_{0}^{-}$ such that the manifolds $\mathcal{N}^{+}\left(m^{+}\right)$
and $\mathcal{N}^{-}\left(m^{-}\right)$ coincide. Consequently, $X_{A}$
vanishes at $m^{+}$ iff it does at $m^{-}$. Following the proof
of Proposition \ref{prop_nombre_A_critic_sur_gamma}, one builds functions
$B_{j}$ which are critical on $\gamma_{0}^{+}$ and automatically
on $\gamma_{0}^{-}$ too.
\end{proof}
Unfortunately, the systems where the periods are equal are non-generic.
We hope that the reader will get convinced by the following example
in dimension $4$ (but can easily be adapted to higher dimensions).

\begin{example}
\label{exp_period_not_equal}Consider the symplectic manifold $\mathcal{M}=\frac{\mathbb{R}}{\mathbb{Z}}\times\mathbb{R}\times\mathbb{R}^{2}$,
with the symplectic form $\omega=d\eta\wedge dt+d\xi\wedge dx$. Let
$F\left(x,\xi\right)=\xi^{2}+\cos x$. Fix a small $\delta>0$. Let
$G\left(x,\xi\right)$ a smooth function compactly supported in $\left\{ x^{2}+\xi^{2}\leq\delta\right\} $.
Assume that $G=c>0$ in the disc $\left\{ x^{2}+\xi^{2}\leq\frac{\delta}{2}\right\} $.
Now, consider the Hamiltonian $H\in C^{\infty}\left(\mathcal{M}\right)$
defined by \[
H\left(t,\eta,x,\xi\right)=\eta\left(1+G\left(x,\xi\right)\right)+F\left(x,\xi\right).\]
First, $H$ is completely integrable since it obviously Poisson-commutes
with $\eta$. One can check that for $\eta$ sufficiently small, the
transversally hyperbolic periodic orbits of $H$ are $\gamma_{\eta,p}\left(t\right)=\left(t,\eta,p,0\right)$,
with $p\in2\pi\mathbb{Z}$. The picture below represents a Poincaré
section ($\eta$ fixed and $t=0$) of the flow of $H$. 

\begin{minipage}[c]{1\columnwidth}%
\begin{center}\includegraphics[%
  scale=0.8]{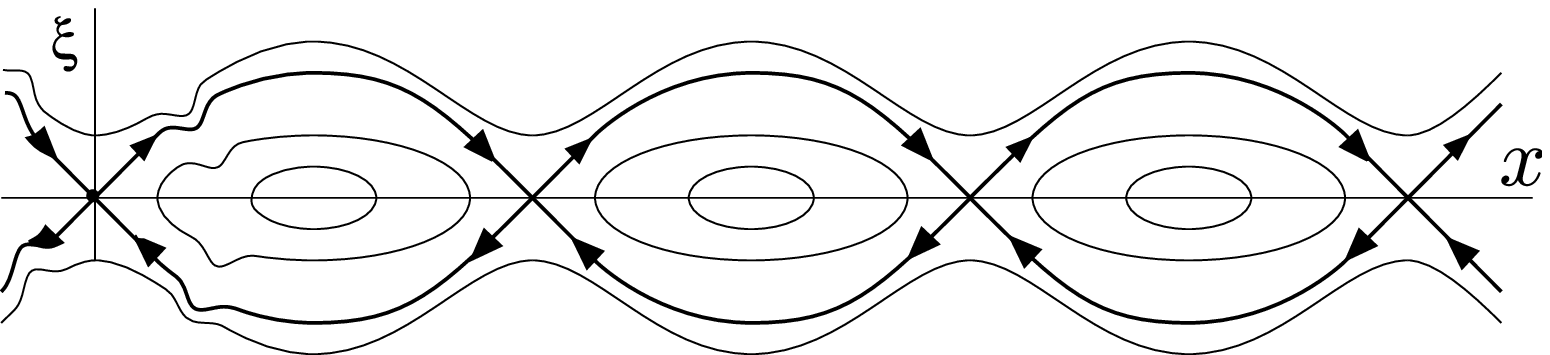}\end{center}\end{minipage}%

Then, a short calculation shows that all the periodic orbits $\gamma_{\eta,p}$
for $p\neq0$ have period $1$ while $\gamma_{\eta,0}$ has period
$\frac{1}{1+c}$. In fact, this example is very general. One can work
on $\mathcal{M}=T^{*}S^{1}\times\mathcal{M}_{0}$, with any symplectic
manifold $\mathcal{M}_{0}$. It is enough to choose a function $F\in C^{\infty}\left(\mathcal{M}_{0}\right)$
with hyperbolic critical points linked by heteroclinic manifolds as
in the picture above, and a function $G\in C^{\infty}\left(\mathcal{M}_{0}\right)$
compactly supported around one of these critical points as above.
\end{example}

\subsubsection{\label{sec_remark_on_potentials}Remark on Mel'nikov potentials}

We would like to conclude this section with a short remark on Mel'nikov
potentials. As mentioned in the introduction, it might happen that
the Mel'nikov $1$-form $\beta$ admits a nice (convergent) integral
expression whereas the Mel'nikov potentials do not (although the potential
itself always exists, as shown in Section \ref{sub_Melnikov_potentials}).
Indeed, suppose that we are in the situation of Proposition \ref{prop_nombre_A_critic_sur_gamma_period_egale},
i.e., the periods on $\gamma_{0}^{+}$ and $\gamma_{0}^{-}$ are equal.
In that case, one has $d$ commuting constants of the motion $A_{1},...,A_{d}$
for which $\beta\left(X_{A_{j}}\right)$ admits the integral expression
given in Theorem \ref{theo_good_A}, namely $A_{1},...,A_{d-1}$ are
those given by Proposition \ref{prop_nombre_A_critic_sur_gamma_period_egale}
and $A_{d}=H_{0}$. Moreover, the associated Hamiltonian vector fields
$X_{j}:=X_{A_{j}}$ provide a global frame on $\mathcal{N}_{0}\setminus\left(\gamma_{0}^{+}\cup\gamma_{0}^{-}\right)$.
Therefore, provided a origin point $m_{0}\in\mathcal{N}_{0}\setminus\left(\gamma_{0}^{+}\cup\gamma_{0}^{-}\right)$
is fixed, one can parameterise%
\footnote{This parameterisation is not injective, but it is surjective.%
} $\mathcal{N}_{0}\setminus\left(\gamma_{0}^{+}\cup\gamma_{0}^{-}\right)$
by $\left(t_{1},...,t_{d}\right)\in\mathbb{R}^{d}\rightarrow m=\phi_{X_{1}}^{t_{1}}\circ...\circ\phi_{X_{d}}^{t_{d}}\left(m_{0}\right)$.
A Mel'nikov potential $L$ is well-defined up to a constant which
can be fixed by setting $L\left(m_{0}\right)=0$. Then, one has \[
L\left(m\right)=\int_{0}^{1}\frac{d}{ds}L\circ\phi_{\sum_{j}t_{j}X_{j}}^{s}\left(m_{0}\right)\, ds=\int_{0}^{1}\sum_{j}t_{j}\beta\left(X_{j}\right)\phi_{\sum_{j}t_{j}X_{j}}^{s}\left(m_{0}\right)ds\]
since the flows of the $X_{j}$'s commute with each other. Inserting
the integral expression of $\beta$ and exchanging the order of the
sum $\Sigma_{j}$ and the integral $\int dt$, one obtains\[
L\left(m\right)=\int_{0}^{1}\left(\int_{-\infty}^{+\infty}\left\{ H_{1},\sum_{j}t_{j}A_{j}\right\} \circ\phi_{X_{H_{0}}}^{t}\circ\phi_{\sum_{j}t_{j}X_{j}}^{s}\left(m_{0}\right)dt\right)ds.\]
Now, since $X_{H_{0}}$ is a symplectic vector field commuting with
the $X_{j}$'s, one has simply \[
\left\{ H_{1},\sum_{j}t_{j}A_{j}\right\} \circ\phi_{X_{H_{0}}}^{t}=\left\{ H_{1}\circ\phi_{X_{H_{0}}}^{t},\sum_{j}t_{j}A_{j}\right\} .\]
Finally, if it was possible to exchange the order of the two integrals
$\int ds$ and $\int dt$, then we would get \[
L\left(m\right)=\int_{-\infty}^{+\infty}\left(-\int_{0}^{1}\frac{d}{ds}H_{1}\circ\phi_{X_{H_{0}}}^{t}\circ\phi_{\sum_{j}t_{j}X_{j}}^{s}\left(m_{0}\right)ds\right)dt.\]
The integration over the $s$ variable would give \[
L\left(m\right)=\int_{-\infty}^{+\infty}\left(H_{1}\circ\phi_{X_{H_{0}}}^{t}\left(m_{0}\right)-H_{1}\circ\phi_{X_{H_{0}}}^{t}\left(m\right)\right)dt.\]
Unfortunately, this integral is not convergent unless we assume that
$H_{1}$ is constant on the orbits $\gamma_{0}^{\pm}$ (in particular
this is the case when $H_{1}$ is critical on $\gamma_{0}^{\pm}$).

\subsection{Recovering the Mel'nikov function}

The Mel'nikov {}``function'' was historically introduced for studying
periodically forced $2$-dimensional Hamiltonian systems. We present
here the class of periodically forced system, as a special example
of the general framework we have been developing throughout this article.

Let $\left(\mathcal{M},\omega\right)$ be a $2d$-dimensional symplectic
manifold and $H_{0}\in C^{\infty}\left(\mathcal{M}\right)$ a Hamiltonian
admitting two fixed points $m_{0}^{+}$ and $m_{0}^{-}$, which are
hyperbolic in the sense that the linear maps which sends $Y\in T_{m_{0}^{\pm}}\mathcal{M}$
to $\left[\tilde{Y},X_{H_{0}}\right]_{m_{0}^{\pm}}$, for any extension
$\tilde{Y}\in\Gamma\left(T\mathcal{M}\right)$, has no eigenvalue
on the imaginary axis. This implies the existence of stable and unstable
manifolds for both points, and we suppose that the stable manifold
$\mathcal{N}_{0}^{+}$ of $m_{0}^{+}$ coincides with the unstable
manifold $\mathcal{N}_{0}^{-}$ of $m_{0}^{-}$. We suppose moreover,
that $H_{0}$ is completely integrable, i.e., there is a momentum
map $\left(A_{1},...,A_{d}\right)$. This hypothesis is automatically
true in the $2$-dimensional case usually considered. Then, we perturb
the Hamiltonian into a $1$-periodic time-dependent Hamiltonian $H_{\varepsilon}\left(t\right)$.
For studying such systems, it is very convenient to consider the {}``extended
system'' on the $\left(2d+2\right)$-dimensional manifold $\tilde{\mathcal{M}}=\mathcal{M}\times T^{*}S^{1}$,
where the $S^{1}$ factor corresponds to the $t$ variable. This manifold
is equipped with the symplectic form $\pi^{*}\omega+d\eta\wedge dt$,
where $\eta$ is the moment variable associated with $t$ and $\pi$
is the projection $\tilde{\mathcal{M}}\rightarrow\mathcal{M}$. Let
us denote the Poisson brackets on $\tilde{\mathcal{M}}$ (resp. $\mathcal{M}$)
by $\left\{ ,\right\} ^{\sim}$ (resp. $\left\{ ,\right\} $). 

The perturbed Hamiltonian $H_{\varepsilon}$ can be viewed as a function
on $\tilde{\mathcal{M}}$ independent on $\eta$. Then, we define
the extended Hamiltonian $\tilde{H}_{\varepsilon}=H_{\varepsilon}\circ\pi+\eta$
and it is easy to check that the dynamics of $H_{\varepsilon}$ is
given by the projection on $\mathcal{M}$ of the dynamics of $\tilde{H}_{\varepsilon}$
on $\tilde{\mathcal{M}}$.

Since the points $m_{0}^{\pm}$ are fixed for $H_{0}$, they give
rise to periodic orbits $t\rightarrow\left(m_{0}^{\pm},\left(\eta,t\right)\right)$
for $\tilde{H}_{\varepsilon}$, denoted by $\gamma_{0}^{\pm}$, and
the hyperbolicity of $m_{0}^{\pm}$ implies the one of $\gamma_{0}^{\pm}$.
The stable manifold of $\gamma_{0}^{+}$, denoted by $\tilde{\mathcal{N}}_{0}^{+}$,
is nothing but the union over all $s\in S^{1}$ of $\phi_{\partial_{t}}^{s}\left(\mathcal{N}_{0}^{+}\right)$,
and coincides with $\tilde{\mathcal{N}}_{0}^{-}$, the unstable manifold
of $\gamma_{0}^{-}$. The pull-back to $\tilde{\mathcal{M}}$ of the
momentum map provides $d$ functions $\tilde{A}_{j}=A_{j}\circ\pi$
which are invariant by the flow of $\tilde{H}_{0}$, since $\left\{ \tilde{H}_{0},\tilde{A}_{j}\right\} _{\left(m,\left(\eta,t\right)\right)}^{\sim}=\left\{ H_{0},A_{j}\right\} _{m}=0$.
Moreover, they are critical on $\gamma_{0}^{\pm}$ since the $A_{j}$'s
are critical on $m_{0}^{\pm}$, because of the hyperbolicity of $m_{0}^{\pm}$.
This means that we can apply Theorem \ref{theo_good_A} which says
that the Mel'nikov $1$-form $\beta$ evaluated on the $X_{\tilde{A}_{j}}$'s
gives 

\[
\beta\left(X_{\tilde{A}_{j}}\right)_{\left(m,\left(\eta,t\right)\right)}=\int_{-\infty}^{+\infty}\left\{ H_{1},\tilde{A}_{j}\right\} ^{\sim}\circ\phi_{X_{\tilde{H}_{0}}}^{s}\left(m,\left(\eta,t\right)\right)ds.\]
Now, the flows of $\tilde{H}_{0}$ and $H_{0}$ are simply related
by \[
\phi_{X_{\tilde{H}_{0}}}^{s}\left(m,\left(\eta,t\right)\right)=\left(\phi_{X_{H_{0}}}^{s}\left(m\right),\left(\eta,t+s\right)\right).\]
 Moreover, since the $\tilde{A}_{j}$'s are pullbacks, then $\left\{ H_{1},\tilde{A}_{j}\right\} _{\left(m,\left(\eta,t\right)\right)}^{\sim}$
is simply equal to $\left\{ H_{1}\left(t\right),A_{j}\right\} _{m}$.
Therefore, the evaluation of $\beta$ becomes \[
\beta\left(X_{\tilde{A}_{j}}\right)_{\left(m,\left(\eta,t\right)\right)}=\int_{-\infty}^{+\infty}\left\{ H_{1}\left(t+s\right),A_{j}\right\} \circ\phi_{X_{H_{0}}}^{s}\left(m\right)\, ds,\]
which is the usual form of the so-called {}``Mel'nikov functions''
$M_{j}\left(t\right)$, for a fixed point $m$. 

\bibliographystyle{plain}
\bibliography{/home/roy/math/biblio/biblio_nico}

\end{document}